\documentclass[a4paper,11pt]{article}
\pdfoutput=1

\usepackage{jcappub}

\usepackage[T1]{fontenc}
\usepackage{multirow}
\usepackage{bm}
\usepackage{mathtools}
\usepackage[version=4]{mhchem}
\usepackage{siunitx}
\usepackage[normalem]{ulem}

\newcommand{\be}{\begin{equation}}
\newcommand{\ee}{\end{equation}}
\newcommand{\de}{{\rm d}}

\compress
\title{Conversations in the dark: cross-correlating birefringence and LSS to constrain axions}

\author[a,b]{S.~Arcari,}
\author[c,d,e]{N.~Bartolo,}
\author[f]{A.~Greco,}
\author[g,h,a]{A.~Gruppuso,}
\author[a,b]{M.~Lattanzi,}
\author[a,b]{and P.~Natoli}
\affiliation[a]{Dipartimento di Fisica e Scienze della Terra, Università degli Studi di Ferrara, Via G. Saragat 1, I-44122 Ferrara, Italy}
\affiliation[b]{Istituto Nazionale di Fisica Nucleare, Sezione di Ferrara, Via G. Saragat 1, I-44122 Ferrara, Italy}
\affiliation[c]{Dipartimento di Fisica e Astronomia “Galileo Galilei”, Universià degli Studi di Padova, Via Marzolo 8, I-35131 Padova, Italy}
\affiliation[d]{Istituto Nazionale di Fisica Nucleare, Sezione di Padova, Via Marzolo 8, I-35131 Padova, Italy}
\affiliation[e]{Istituto Nazionale di Astrofisica - Osservatorio Astronomico di Padova, Vicolo dell’Osservatorio 5, I-35122 Padova, Italy}
\affiliation[f]{Department of Astronomy, University of Florida,\\211 Bryant Space Science Center, Gainesville, FL 32611, USA}
\affiliation[g]{Istituto Nazionale di Astrofisica - Osservatorio di Astrofisica e Scienza dello Spazio di Bologna, Via Gobetti 101, I-40129 Bologna, Italy}
\affiliation[h]{Istituto Nazionale di Fisica Nucleare, Sezione di Bologna, Viale Berti Pichat 6/2, I-40127 Bologna, Italy}

\emailAdd{stefano.arcari@unife.it}
\emailAdd{nicola.bartolo@pd.infn.it}
\emailAdd{alessandro.greco@ufl.edu}
\emailAdd{alessandro.gruppuso@inaf.it}
\emailAdd{massimiliano.lattanzi@fe.infn.it}
\emailAdd{paolo.natoli@unife.it}

\abstract{Unveiling the dark sector of the Universe is one of the leading efforts in theoretical physics. Among the many models proposed, axions and axion-like particles stand out due to their solid theoretical foundation, capacity to contribute significantly to both dark matter and dark energy, and potential to address the small-scale crisis of $\Lambda$CDM. Moreover, these pseudo-scalar fields couple to the electromagnetic sector through a Chern-Simons parity-violating term, leading to a rotation of the plane of linearly polarized waves, namely cosmic birefringence. We explore the impact of the axion-parameters on anisotropic birefringence and study, for the first time, its cross-correlation with the spatial distribution of galaxies, focusing on ultralight axions with masses $10^{-33}\,\si{\electronvolt}\le m_\phi\le10^{-28}\,\si{\electronvolt}$. Through this novel approach, we investigate the axion-parameter space in the mass $m_\phi$ and initial misalignment angle $\theta_i$, within the framework of early dark energy models, and constrain the axion-photon coupling $g_{\phi\gamma}$ required to achieve unity in the signal-to-noise ratio of the underlying cross-correlation, computed with the instrument specifications of \emph{Euclid} and forthcoming CMB-polarization data. Our findings reveal that for masses below $10^{-32}\,\si{\electronvolt}$ and initial misalignment angles greater in absolute value than $\pi/4$, the signal-to-noise ratio not only exceeds unity but also surpasses that achievable from the auto-correlation of birefringence alone (up to a factor 7), highlighting the informative potential of this new probe. Additionally, given the late-time evolution of these low-mass axions, the signal stems from the epoch of reionization, providing an excellent tool to single out the birefringence generated during this period.}

\begin{document}
\maketitle

\section{Introduction}
\label{sec:intro}

The Cosmic Microwave Background (CMB) stands as one of the most invaluable windows into the early universe, offering profound insights into the fundamental nature of cosmology. Over the years, successive data releases from the {\it Planck} satellite \citep{Planck:2013pxb,Planck:2015fie,Planck:2018vyg, Planck:2018nkj} have provided an unprecedented level of precision in mapping the temperature anisotropies of the CMB. These measurements have not only affirmed the predictions of the standard cosmological model but have also unveiled subtle discrepancies that challenge our current understanding. In this pursuit, the polarization data of the CMB has emerged as a pivotal tool, offering unique insights into the primordial universe. By probing the polarization patterns embedded within the CMB, researchers can delve deeper into the dynamics of cosmic inflation, test fundamental physics principles such as parity violation, and unravel the intricate tapestry of the early universe. In this context, a prominent consequence of violating parity is the in vacuo rotation of the linear polarization plane of CMB radiation, known as the cosmic birefringence (CB) effect\footnote{It's worth noting that birefringence represents just one potential outcome of parity-violating physics. Such physics can lead to a mixing of the CMB polarization modes, resulting in a non-zero EB correlation. Some models explore the possibility of explaining this correlation through, e.g., primordial chiral gravitational waves \citep{Alexander:2004wk,Lyth:2005jf,Alexander:2009tp,Dyda:2012rj,Gerbino:2016mqb,Fujita:2022qlk,Bartolo:2017szm,Bartolo:2018elp,Bartolo:2020gsh}, without causing any rotation in the plane of linearly polarized waves.}. Specifically, cosmic birefringence can be sourced by Chern-Simons terms in the electromagnetic Lagrangian \citep{Carroll:1989vb, Carroll:1998zi, Li:2008tma, Pospelov:2008gg, Finelli:2008jv}, including quintessence couplings to the electromagnetic field \citep{Balaji:2003sw, Liu:2016dcg, Capparelli:2019rtn, Caldwell:2011pu, Galaverni:2014gca, Komatsu:2022nvu}, by quantum-gravity motivated effective theories for electromagnetism \citep{Myers:2003fd, Gubitosi:2009eu, Gubitosi:2010dj}, such as Lorentz-violating electrodynamics \citep{Kostelecky:2007zz,Caloni:2022kwp}, or by astrophysical sources \citep{Carroll:1989vb,Galaverni:2014gca} (with a frequency-dependent effect). Quintessence, a hypothetical field permeating the Universe, may be accounted for by various mechanisms. These include minimally coupled scalar fields like axion-like particles (ALPs) \citep{Preskill:1982cy,Abbott:1982af,Dine:1982ah,Liu:2016dcg,Nakagawa:2021nme,Obata:2021nql,Arvanitaki:2009fg,Hlozek:2014lca,Marsh:2015xka,Poulin:2018dzj,Kim:2021eye,Jain:2021shf,Lin:2022niw,Greco:2022xwj, Gonzalez:2022mcx, Gasparotto:2023psh,Greco:2024oie} and early dark energy (EDE) from the string axiverse \citep{Poulin:2018cxd,Capparelli:2019rtn,Fujita:2020ecn,Choi:2021aze,Murai:2022zur,Gasparotto:2022uqo,Kamionkowski:2022pkx}, or non-minimally coupled scalar fields found in scalar-tensor theories \citep{Ballardini:2020iws,Ballardini:2023mzm} and early modified gravity \citep{Braglia:2020auw}.
Analysis techniques to measure the CB effect in CMB polarization data have been proposed and used in refs.~\citep{Namikawa:2021gbr, Minami:2020odp, Komatsu:2022nvu, Fujita:2022qlk, Diego-Palazuelos:2022dsq, Eskilt:2022wav} to obtain measurements of a non-vanishing isotropic CB angle, up to a 3-$\sigma$ significance, both with the {\it Planck} PR3 \citep{Planck:2018vyg} and PR4 \citep{Planck:2020olo} releases. Additionally, increasing interest is dedicated to investigating the variation of this effect across the sky, namely anisotropic cosmic birefringence, through data from {\it Planck} \citep{Gruppuso:2020kfy, Bortolami:2022whx}, BICEP/KECK \citep{BICEPKeck:2022kci} and SPT \citep{SPT:2020cxx}. Cross-correlations between anisotropic cosmic birefringence and both CMB temperature and polarization have been studied in refs.~\citep{Caldwell:2011pu, Capparelli:2019rtn, Greco:2022ufo,Greco:2022xwj} as a tool for investigating the properties of EDE and axion models. 

In this work, we expand upon previous studies by examining, for the first time, the cross-correlation between ALP-induced anisotropic birefringence and galaxy number counts. We compute this signal with a targeted modification of the Boltzmann code {\tt CLASS} \citep{DiDio:2013bqa,Lesgourgues:2011re}. We show how this new probe can be particularly sensitive to ultralight ALPs , given that their late-time evolution aligns with the epoch of interest for the spatial distribution of galaxies. This correlation stems from a rotation of linearly polarized waves induced at the epoch of reionization, underscoring the novelty and worth of our approach in untangling the origin of birefringence. We focus on axions with masses $m_\phi=\left[10^{-33},\;10^{-28}\right]\,\si{\electronvolt}$, whose dynamic evolution initiates after the epoch of recombination. Within this domain, the lightest candidates experience significantly higher signal-to-noise ratios ($S/N$), even exceeding unity, with respect to larger masses, stemming from the combination of forthcoming CMB data and galaxy surveys. With this in mind, we explore the ALP-parameter space, revealing a region where not only the $(S/N)$ of the cross-correlation exceeds one but also surpasses that achieved by the sole cosmic birefringence auto-correlation.\footnote{Moreover, this constrained region is in agreement, to a large extent, with that able to reproduce the isotropic birefringence angle as measured by recent studies \citep{Greco:2024oie}.} The effect is maximal around $m_\phi\sim10^{-32}\,\si{\electronvolt}$ and high initial misalignment angles $\theta_i$, with the cross-correlation being up to a factor 7 more informative than the auto-correlation. In conclusion, the innovative approach presented in this study not only offers a promising method for constraining axion-like particles and measuring birefringence but also has the potential to enhance our understanding of fundamental physics and cosmological phenomena.

The paper is organized as follows: in section \ref{sec:theory} we summarize the theoretical framework behind ALP-induced cosmic birefringence, with particular attention to the underlying axion-like model and its field dynamics. Section \ref{sec:cross} goes through the formalism to derive the cross-correlation angular power spectrum of birefringence and galaxies and the related variance given by forthcoming CMB experiments and galaxy surveys, together with a discussion about important axion-parameters. In section \ref{sec:res} we show our results and finally conclusions are drawn in section \ref{sec:conc}. Appendix \ref{app:A} serves as an informative summary of possible implications on the results of changing the ALP potential, whilst appendix \ref{app:B} discusses the impact of including redshift tomography in the analysis.

Throughout the paper we assume a spatially flat $\Lambda$CDM cosmology with cosmological parameters as derived by the {\it Planck} satellite in 2018 \citep{Planck:2018vyg}, otherwise specified\footnote{Let us keep in mind that one of the main ingredients of our work, ALPs, are of course an extension of the underlying cosmological model.}.

\section{The theoretical framework of ALP-induced cosmic birefringence}
\label{sec:theory}
Parity-violating terms in the electromagnetic Lagrangian give rise to different phase velocities of right- and left-handed states of photons, leading to a rotation of the plane of linear polarization on the sky (known as cosmic birefringence). Notably, Chern-Simons terms \citep{Carroll:1989vb, Carroll:1998zi} have been extensively investigated in this context. Here, an additional term in the electromagnetic Lagrangian is given by a pseudo-scalar field $\phi$ coupled to photons:
\be\label{eq:lagrangian}
    \mathcal{L} \supset -\frac{1}{2}\,g^{\mu\nu}\partial_\mu\phi\,\partial_\nu\phi - V(\phi) -\frac{1}{4}\,g_{\phi\gamma}\,\phi\,F_{\mu\nu}\tilde F^{\mu\nu} \;,
\ee
where $g^{\mu\nu}$ is the metric tensor, $V(\phi)$ the axion potential, $g_{\phi\gamma}$ the field-to-photon coupling (later referred as axion-photon coupling, for the purpose of our work), and $\tilde F^{\mu\nu} = \epsilon^{\mu\nu\rho\sigma}F_{\rho\sigma}/2$ the Hodge dual of the Maxwell tensor $F_{\mu\nu}$, with $\epsilon^{\mu\nu\rho\sigma}$ being the Levi-Civita antisymmetric symbol. The field $\phi$ must retain a pseudo-scalar nature to preserve parity-invariance within the Lagrangian density, as the term $F_{\mu\nu}\tilde F^{\mu\nu}$ violates parity. Despite the field choice can be nearly arbitrary, axion-like fields emerge as compelling candidates \citep{Gasparotto:2022uqo,Nakagawa:2021nme,Choi:2021aze,Takahashi:2020tqv,Preskill:1982cy,Obata:2021nql,Marsh:2015xka,Poulin:2018dzj,Jain:2021shf,Agrawal:2022lsp}. Indeed, interest in axion and axion-like physics has surged due to their manifold implications in string theories \citep{Witten:1984dg,Conlon:2006tq,Svrcek:2006yi,Vilenkin:1982ks,Huang:1985tt,Davis:1986xc,Arvanitaki:2009fg, Marsh:2015xka, Kitajima:2022jzz, Jain:2021shf}, dark matter physics \citep{Abbott:1982af,Lin:2022niw,Liu:2016dcg}, quintessence models \citep{Fujita:2020ecn,Caldwell:2011pu}, and possible solutions of the Hubble tension, through EDE models \citep{Kamionkowski:2022pkx,Poulin:2018cxd,Poulin:2018dzj,Capparelli:2019rtn, Efstathiou:2023fbn}.

The main consequence of eq.~\eqref{eq:lagrangian} is a modification of Maxwell field equations, which admit plane wave solutions for the electric and magnetic field that correspond to left and right circularly polarized waves to lowest order, whose superposition is a linearly polarized wave \citep{Harari:1992ea}. However, due to the Chern-Simons' interaction they propagate at different speed, getting out of phase or, equivalently, inducing a rotation of the linear polarization in question (i.e. CMB). The angle of rotation along the pathway of CMB is given by
\be\label{eq:alpha_inst}
    \alpha({\bf\hat{n}}) = \frac{1}{2}\,g_{\phi\gamma}\int_{\tau_s}^{\tau_0}\de\tau\;\left(\frac{\partial}{\partial\tau}-{\bf\hat{n}\cdot\nabla}\right)\phi(\tau,\;\Delta\tau\,{\bf\hat{n}}) = \frac{1}{2}\,g_{\phi\gamma}\left[\phi(\tau_0,\;{\bf0})-\phi(\tau_s,\;\Delta\tau_s\,{\bf\hat{n}})\right] \;,
\ee
with $\tau_0$ and $\tau_s$ being the conformal time today and at recombination, respectively. The coming direction of the electromagnetic wave is $-{\bf \hat n}$, through wich we can define the starting point of the photon’s path ${\bf x}=\Delta\tau\,{\bf\hat{n}}=(\tau_0-\tau)\,{\bf\hat{n}}$. Eq.~\eqref{eq:alpha_inst} holds in the sudden recombination approximation, i.e., assuming that all photons are emitted instantaneously at the last scattering surface. However, photons of the CMB are not actually emitted all at the same time, but statistically distributed over the photon visibility function $g(\tau)$ \citep{Capparelli:2019rtn,Greco:2024oie,Liu:2006uh,Kosowsky:1996yc,Finelli:2008jv,Gubitosi:2009eu,Galaverni:2023zhv}:
\be\label{eq:alpha}
    \alpha({\bf\hat{n}}) = \frac{1}{2}\,g_{\phi\gamma}\,\phi(\tau_0,\;{\bf 0})-\frac{1}{2}\,g_{\phi\gamma}\int_0^{\tau_0}\de\tau\;g(\tau)\,\phi(\tau,\;\Delta\tau\,{\bf\hat{n}}) \;.
\ee
This equation reduces to Eq.~\eqref{eq:alpha_inst} when the visibility function is approximated as a delta function centered in $\tau_s$. Eqs.~\eqref{eq:alpha_inst} and \eqref{eq:alpha} show that birefringence is a propagation effect, hence, the longer the path, possibly the larger the effect. For this reason, CMB offers promising observational prospects. Cosmic birefringence violates parity symmetry\footnote{While the Lagrangian of eq.~\eqref{eq:lagrangian}, including the Chern-Simons term is parity-invariant, the time-dependence of the pseudoscalar field $\phi$ introduces an effective parity violation.} leading to a correlation between E and B modes in the CMB \citep{Lue:1998mq}. This effect can be deeply constrained at small angular scales by forthcoming ground-based experiment, such as Simons Observatory (SO) \citep{SimonsObservatory:2018koc}, South Pole Observatory \citep{Moncelsi:2020ppj} and CMB-S4 \citep{CMB-S4:2020lpa}. Meanwhile, larger scales can be precisely measured by LiteBIRD (LB) \citep{LiteBIRD:2022cnt}. The target observational-scale of birefringence is strongly dependent on the properties of the underlying pseudo-scalar and will be extensively discussed in the following.

Anisotropies in the rotation angle across the sky can stem from the fluctuations of the scalar field. To this end, it is useful to exploit a perturbative approach to study the birefringence effect in its isotropic (induced by the background field $\bar\phi$) and anisotropic (induced by the field perturbation $\delta\phi$) part.  Over recent years, significant observational constraints on isotropic birefringence have been established using subsequent data releases of the {\it Planck} satellite \citep{Planck:2016soo,Minami:2020odp,Diego-Palazuelos:2022dsq,Eskilt:2022cff}, and further enhanced through a combination of WMAP and {\it Planck} datasets \citep{Eskilt:2022wav}. Concurrently, investigations into anisotropic birefringence have been conducted using various datasets (including {\it Planck} \citep{Gruppuso:2020kfy, Bortolami:2022whx}, BICEP/KECK \citep{BICEPKeck:2022kci} and SPT \citep{SPT:2020cxx}) alongside theoretical explorations involving cross-correlations with CMB in both axion-like physics scenarios \citep{Zhai:2020vob, Greco:2022xwj, Greco:2024oie, Greco:2022ufo} and EDE models \citep{Caldwell:2011pu, Capparelli:2019rtn, Eskilt:2023nxm}.

\subsection{Background evolution}
\label{subsec:bkg}
Including the Chern-Simons term of eq.~\eqref{eq:lagrangian}, we derive the dynamics of the background field $\bar\phi$ via its equation of motion (EoM)
\be\label{eq:EOMbkg}
    \bar{\phi}''+2\mathcal{H}\bar{\phi}'+a^2\frac{\de V}{\de\bar{\phi}}=0 \,,
\ee
where $a$ is the scale factor, $\mathcal{H}$ the conformal Hubble parameter and primes denote derivatives with respect to conformal time $\tau$. The evolution is governed by the choice of the potential $V$ and its interplay with the Hubble expansion rate. In this work, we focus on EDE models for ALPs, as discussed in section \ref{subsec:potential}. However, for illustrative purposes, let us consider a quadratic potential $V=m_\phi^2\phi^2/2$ (resulting in $a^2\,m^2_\phi\phi$ as last term of eq.~\eqref{eq:EOMbkg}). Assuming the field to be locked at early times ($\bar\phi'=0$), it becomes evident from eq.~\eqref{eq:EOMbkg} that it stays so as long as the Hubble expansion rate significantly exceeds the field's mass. Hence, lying within a slow-roll phase before transitioning to oscillations when $2\mathcal{H}\sim a\,m_\phi$. Notably, if $m_\phi>10^{-28}\,\si{\electronvolt}$, oscillations commence before the recombination epoch, whereas for smaller masses, the field evolves at a later stage.\footnote{Let us notice that for $m_\phi\sim10^{-32}\,\si{\electronvolt}$, the evolution starts around the epoch of reionization, which aligns to the region interested by our findings (see section \ref{sec:res}).} Additionally, for $m_\phi<10^{-33}\,\si{\electronvolt}$, the field's evolution has not initiated, as $H_0\sim10^{-33}\,\si{\electronvolt}$. 

In the sudden recombination approximation, the isotropic birefringence effect is solely determined by the difference in the field between recombination and the present time (as depicted in eq.~\eqref{eq:alpha_inst}). Fields that start evolving sensibly before recombination (i.e. $m_\phi>10^{-28}\,\si{\electronvolt}$) would produce a fainter birefringence signal due to the relatively small difference between these two periods.  Similarly, very light scalars ($m_\phi\sim10^{-33}\,\si{\electronvolt}$) would exhibit a comparable behavior since their evolution at very late times also results in a small difference. Hence, we expect intermediate mass ranges to give rise to the largest effect (see ref.~\citep{Greco:2022xwj,Nakatsuka:2022epj, Greco:2022ufo} for a detailed discussion). The mass of the scalar also governs its phenomenological behavior: if oscillations commence early enough they can contribute to sourcing matter perturbations, effectively behaving as dark matter; otherwise, a late evolution of the field can trigger an accelerated expansion of the Universe, mimicking the characteristics of dark energy. Achieving an accurate estimation of the critical mass, which separates these two regimes, necessitates a deep study of the field’s equation of state. By varying the scalar sector of the Lagrangian, with respect to the metric tensor $g_{\mu\nu}$, we can derive the energy-momentum tensor of the field
\be
    T_{\mu\nu} = \partial_\mu\phi\,\partial_\nu\phi-g_{\mu\nu}\left(\frac{1}{2}g^{\sigma\rho}\,\partial_\sigma\phi\,\partial_\rho\phi+V(\phi)\right) \;,
\ee
and express its background density and pressure
\be\label{eq:energy}
    \begin{split}
        \rho_{\bar\phi}&=-{T^0}_0=\frac{1}{2a^2}\bar\phi'^2+V(\phi)\\
        P_{\bar\phi}&=\frac{1}{3}{\delta^i}_j{T^j}_i=\frac{1}{2a^2}\bar\phi'^2-V(\phi) \;.
    \end{split}
\ee
Consequently, the equation of state reads
\be\label{eq:EoSbkg}
    w_{\bar\phi}=\frac{P_{\bar\phi}}{\rho_{\bar\phi}}=\frac{\bar\phi'^2-2a^2\,V(\phi)}{\bar\phi'^2+2a^2\,V(\phi)} \;.
\ee
During its initial slow roll phase, characterized by $\bar\phi'\sim0$, the field exhibits properties akin to dark energy and maintains this behavior as long as $w_{\bar\phi}<-1/3$. Subsequently, for large enough masses, it transitions to an oscillatory state, resembling a pressure-less fluid, reminiscent of dark matter. As noted in previous works \citep{Greco:2022xwj,Nakatsuka:2022epj}, fields with masses lower than the current Hubble rate ($m_\phi\lesssim10^{-33}\,\si{\electronvolt}$) continue to manifest characteristics akin to dark energy today, with a tendency towards larger equations of state for higher masses. Instead, for masses exceeding $m_\phi\gtrsim10^{-31}\,\si{\electronvolt}$, these fields behave akin to dark matter in the present epoch, even though they may have contributed to dark energy in earlier stages. This underscores the profound interest in exploring the parameter space of mass, offering insights into both potential ramifications on observations of birefringence and a comprehensive exploration of its phenomenological diversity.

Linear polarization is produced not only within a relatively narrow time-window at last scattering but also during reionization \citep{Zaldarriaga:1996ke}. A detailed discussion on the birefringence effect generated by this two distinct epochs can be found in refs.~\citep{Greco:2022xwj,Nakatsuka:2022epj}, nonetheless, in our work we will always consider the full visibility function, as our purpose is to constrain the pseudo-scalar parameter space rather than exploring the epoch-dependence of cosmic birefringence.

Finally, let us note that, for the whole discussion, we will set ourselves in the "spectator field approximation", assuming that the background field $\bar\phi$ does not contribute to Einstein equations (i.e. has no impact on the Hubble expansion rate and its energy budget is negligible with respect to the total energy of the Universe).\footnote{It is evident how this regime does not allow modifications of the Hubble parameter today, and consequent alleviations of the Hubble tension as discussed in ref.~\citep{Poulin:2018cxd}. Our goal is to provide a novel probe of birefringence and explore its detectability in future surveys. Nonetheless, we briefly illustrate the behavior of the axion energy density across the whole parameter space in appendix~\ref{app:C}.}

\subsection{Perturbation evolution}
\label{subsec:prt}
As previously mentioned, anisotropies in the rotation angle of eq.~\eqref{eq:alpha} can be studied by separating the background field $\bar\phi$ (only time-dependent) and the perturbation field $\delta\phi$ (both time- and space-dependent),
\be
    \phi(\tau,\;{\bf x})=\bar\phi(\tau)+\delta\phi(\tau,\;{\bf x}) \;.
\ee
Again, by varying the kinematic action, the EoM of the field perturbation, expanded in Fourier space, reads
\be\label{eq:EoMpert}       
    \delta\phi''+2\mathcal{H}\,\delta\phi'+\left(k^2+a^2\frac{\de^2V}{\de\bar{\phi}^2}\right)\delta\phi=-\frac{1}{2}h'\bar{\phi}' \;,
\ee
where $h$ is the metric perturbation in the synchronous gauge \citep{Lifshitz:1945du,Ma:1995ey}. Following the literature \citep{DiDio:2013bqa,Greco:2022xwj,Greco:2024oie}, we solve eq.~\eqref{eq:EoMpert} with initial perturbations and related derivative set to zero ($\delta\phi_\text{ini}=0$, $\delta\phi'_\text{ini}=0$), exploiting the attractor to adiabatic initial conditions incorporated in {\tt CLASS}. The field starts to exhibit anisotropies when the background's oscillations and its potential inject power into the perturbations. The source term of eq. \eqref{eq:EoMpert} remains null as long as the background field is locked (i.e. $\bar{\phi}'=0$), hence, we anticipate perturbations to commence their evolution around the same time. Needless to say that the perturbation's evolution strongly depends on the nature of the field's potential, which governs the underlying phenomenology\footnote{There exist a certain class of potentials for which the perturbative approach does not hold at every Fourier scale, and could potentially be interesting to be investigated. A more detailed discussion about it can be found in appendix \ref{app:A}.}.

Introducing the perturbation of the pseudo-scalar in eq.~\eqref{eq:alpha} leads to anisotropic birefringence
\be
    \alpha({\bf \hat n}) = \bar\alpha+\delta\alpha({\bf \hat n}) \;,
\ee
where $\bar\alpha$ is the isotropic counterpart (i.e., dependent on the background field, which does not vary with the direction on the sky) and $\delta\alpha({\bf \hat n})$ can be expressed by expanding the perturbation-dependent part of eq.~\eqref{eq:alpha} in spherical harmonics
\be
    \delta\alpha({\bf \hat n}) = \sum_{\ell m}\alpha_{\ell m}Y_{\ell m}({\bf \hat n}) \;.
\ee
A further expansion in Fourier space yields (see also refs.~\citep{Zhao:2014yna,Li:2008tma})
\be\label{eq:alm}
    \alpha_{\ell m} = \frac{i^\ell g_{\phi\gamma}}{(2\pi)^2}\int_0^{\tau_0}\de\tau\;g(\tau)\,\int {\rm d}^3{\bf k}\;Y_{\ell m}^*({\bf\hat k})\,\delta\phi(\tau,\;{\bf k})j_\ell\left[k(\tau_0-\tau)\right] \,,
\ee
$j_\ell\left[k(\tau_0-\tau)\right]$ being the $\ell$-th spherical Bessel function. Having this information at our disposal, deriving the related two-point statistics is straightforward, and we postpone this task to section~\ref{sec:cross}.

\subsection{The ALP potential}
\label{subsec:potential}
Peccei and Quinn proposed a dynamical mechanism to solve the strong CP problem \citep{Peccei:1977hh,Peccei:2006as}, through a new Abelian $U(1)_{PQ}$ symmetry, which is spontaneously broken below an energy scale $f_a$, generating a pseudo Nambu-Goldstone boson, namely the QCD axion \citep{Weinberg:1977ma}. The latter could have been produced in the early Universe both through non-thermal processes (such as the decay of topological defects \citep{Abbott:1982af,Preskill:1982cy,Dine:1982ah} and vacuum realignment \citep{Huang:1985tt,Vilenkin:1982ks,Davis:1986xc}) and thermal processes (including scattering in the standard model plasma \citep{Turner:1986tb,Salvio:2013iaa} and the decay product of a parent particle \citep{Conlon:2013isa,Higaki:2013lra,Cicoli:2012aq}). Nonetheless, extensions of the standard model can lead to light particles with similar properties, i.e. axion-like particles. Among the possible production mechanisms, we have string theory compactification \citep{Svrcek:2006yi,Vilenkin:1982ks,Huang:1985tt,Davis:1986xc,Arvanitaki:2009fg, Hlozek:2014lca, Marsh:2015xka, Kitajima:2022jzz, Jain:2021shf}, spontaneous breaking of global symmetries \citep{McDonald:1993ex,Lin:2022niw, Cacciapaglia:2019bqz,Burgess:2000yq}, or accidental symmetries \citep{Harigaya:2013vja,DiLuzio:2017tjx}. This axion-like component can lead to various gravitational and electromagnetic signatures, and provide for the whole or part of the present-day dark matter or dark energy density.

In this work, we consider an EDE model from the string axiverse \citep{Svrcek:2006yi,Arvanitaki:2009fg,Kamionkowski:2014zda} with a cosine-like potential arising non-perturbatively,
\be\label{eq:potential}
    V(\phi) = m_\phi^2\,f_a^2\left[1-\cos{\frac{\phi}{f_a}}\right]^n \,,
\ee
where $f_a$ is the scale of the spontaneous breakdown of a continuous symmetry, generating the pseudo Nambu-Goldstone boson, as a spin-0 degree of freedom. This potential is ideal both for dark energy and dark matter, given its residual shift-symmetry $\phi\rightarrow\phi+\text{const}$ and oscillatory behavior. At symmetry breaking the field is overdamped, being frozen by Hubble friction. Here, the axion field undergoes a coherent initial displacement, namely the initial misalignment $\theta_i=\phi_i/f_a$, that must be phase-confined between $-\pi$ and $\pi$ for the QCD axion, owing to the construction of the potential. This does not hold for ALPs, that could potentially inhabit a distinct vacuum state, under shift symmetry. Nonetheless, additional interactions with the standard model could emerge in this case, yet we opt to adhere to the conventional parameter space ($-\pi<\theta_i<\pi$). Furthermore, we fix $f_a$ to the {\it Planck} mass $M_\text{Pl}$, and consider the field's mass $m_\phi$, the initial misalignment angle $\theta_i$ and, of course, the axion-photon coupling $g_{\phi\gamma}$ as free parameters of our analysis. For the latter, we expect a somewhat proportionality to $1/f_a$, however, we can always fix the decay constant and remap the mass ($m_\phi\rightarrow m_\phi\,f_a$) and the initial misalignment ($\theta_i\rightarrow\theta_i/f_a$) to retain the coupling as a free, independent parameter.

When the Hubble rate drops below the effective mass of the field, the latter begins to oscillate around the minimum of the potential. Fast oscillations produce an effect similar to that of a fluid with averaged EoS $\langle w \rangle = (n-1)/(n+1)$ \citep{Turner:1983he,Masso:2005zg}. For $n=1$, we recover the classical axion potential \citep{Marsh:2010wq}, with a cold dark matter behavior around the potential's minimum ($\propto\phi^2$), since $\langle w \rangle=0$. Whereas, $n=2$ leads to a radiation-like component ($\langle w \rangle=1/3$) \citep{Greco:2022xwj,Capparelli:2019rtn}, and $n>2$ to a fluid that dilutes faster than radiation. In particular, the case $n=3$ has proved extremely interesting in the context of alleviating the Hubble tension \citep{Capparelli:2019rtn} and reproducing the current constraint on isotropic birefringence \citep{Greco:2024oie}\footnote{It's important to emphasize that this case study presents a component that dilutes more rapidly than any other standard components of the Universe, thereby having no impact on its late evolution.}. Moreover, for $n\le2$, the axion-like field ceases to be treatable~perturbatively at some scales and different approaches should be taken into consideration. The case $n=2$ is particularly cumbersome, as it reflects onto a divergent manifestation of Floquet theory \citep{ASENS_1883_2_12__47_0}, leading to an exponentially growing solution of the EoM \citep{Amin:2011hu,Smith:2023fob}. For these reasons, we will fix the exponent to 3 in the following and briefly discuss different cases in appendix \ref{app:A}.

\section{The cross-correlation signal of birefringence and galaxies}
\label{sec:cross}
The ALPs' parameter space has been constrained in ref.~\citep{Fujita:2020ecn, Greco:2024oie}, exploiting the isotropic birefringence angle obtained by \citep{Minami:2020odp}, with the {\it Planck} 2018 polarization data \citep{Planck:2018vyg}. Moreover, its cross-correlation with CMB anisotropies has proved as a useful tool to constrain the underlined ALP model \citep{Greco:2024oie, Capparelli:2019rtn}. At the same time, galaxies are also powerful probes of axion-like physics \citep{Lague:2021frh,Diehl:2021gna}, especially in view of the forthcoming injection of new data from the recently launched \emph{Euclid} survey \citep{Euclid:2021icp}, as well as DESI~\citep{DESI:2013agm,DESI:2016fyo} and SPHEREx~\citep{SPHEREx:2014bgr}. In this work, we cross-correlate, for the first time, the information coming from cosmic birefringence with that of galaxy counts. We propose this angular cross-correlation as a novel way of constraining ALPs:
\be\label{eq:alphaG}
    C_\ell^{\alpha G_i} = 4\pi\int \frac{{\rm d}k}{k}\;\mathcal{P}_\mathcal{R}(k)\Delta_\ell^\alpha(k)\Delta_\ell^{G_i}(k) \,,
\ee
where $\mathcal{P}_\mathcal{R}$ is the primordial power spectrum, and $\Delta_\ell^\alpha(k)$ is the birefringence kernel obtained from the angular coefficients $\alpha_{\ell m}$ of eq.~\eqref{eq:alm} convolved with the photon visibility function $g(\tau)$ over the line of sight. This reads
\be\label{eq:alphakernel}
     \Delta^\alpha_\ell(k) = g_{\phi\gamma}\int_0^{\tau_0}{\rm d}\tau\;g(\tau)T_{\delta\phi}(\tau,\,k)j_\ell\left[k(\tau_0-\tau)\right] \,.
\ee
Let us underline the physical richness of such a correlation, as both probes (birefringence anisotropies and galaxy overdensities) are sourced by scalar perturbations in the Einstein-Boltzmann equations \citep{Li:2008tma,Greco:2022xwj,Greco:2022ufo}. In eq.~\eqref{eq:alphakernel}, $T_{\delta\phi}(\tau,\,k)$ is a transfer function between the primordial power spectrum $\mathcal{P}_\mathcal{R}(k)$ and that of the field fluctuations $\delta\phi$, obtained from the latter dynamics\footnote{By looking at the right hand side of eq.~\eqref{eq:EoMpert}, it becomes evident how the evolution of axion perturbations is directly sourced by gravitational potentials and thereby deeply connected to the primordial scalar perturbations.}, and enters the two-point cross-correlation function
\be
    \langle\delta\phi(\tau,\;{\bf k})\delta\phi(\tau,\;{\bf k'})\rangle=\frac{2\pi^2}{k^3}\mathcal{P}_\mathcal{R}(k)T_{\delta\phi}(\tau, k)T_{\delta\phi}(\tau, k')(2\pi)^3\delta^{(3)}({\bf k}-{\bf k}') \,.
\ee
In eq.~\eqref{eq:alphaG}, $\Delta_\ell^{G_i}(k)$ is the galaxy kernel in the $i$-th redshift bin and reads
\be\label{eq:galkernel}
    \Delta^{G_i}_\ell = \int{\rm d}z\;\frac{{\rm d}N}{{\rm d}z}W_i(z)\Delta_\ell(z,\,k) \,,
\ee
where ${\rm d}N(z)/{\rm d}z$ is the galaxy redshift distribution and $W_i(z)$ the selection function of the $i-$th redshift bin. While the analysis in the main text is performed considering a single redshift bin, a comparison with a tomographic approach, exploiting \emph{Euclid}~\citep{Euclid:2019clj} standard prescription for the redshift bins and selection functions $W_i(z)$, can be found in appendix~\ref{app:B}. Ultimately, $\Delta_\ell(z,\,k)$ is the galaxy kernel \citep{DiDio:2013bqa,Challinor:2011bk,Bonvin:2011bg}
\begin{table}[t]
\centering
\begin{tabular}{l|ccc}
\hline\rule{0mm}{5mm}
~& $\sigma_P$ [$\mu$K arcmin] & $\theta_{\rm fwhm}$ [arcmin] & $f_{\rm sky}$\\[1mm]
\hline
\hline\rule{0mm}{5mm}LiteBIRD & $3.11$ & $30$ & $0.7$ \\[1mm]
Simons Observatory LAT & $8.49$ & $1.4$ & $0.6$ \\[1mm]
CMB-S4 & $4.24$ & $1$ & $0.6$ \\[1mm]
\hline
\end{tabular}
\caption{Forecast sensitivity on CMB polarization ($\sigma_P$), angular resolution ($\theta_{\rm fwhm}$) and observed fraction of the sky ($f_{\rm sky}$) for future CMB experiments.}
\label{tab:noise}
\end{table}
\begin{figure}[t]
\centering
\includegraphics[width=0.95\hsize]{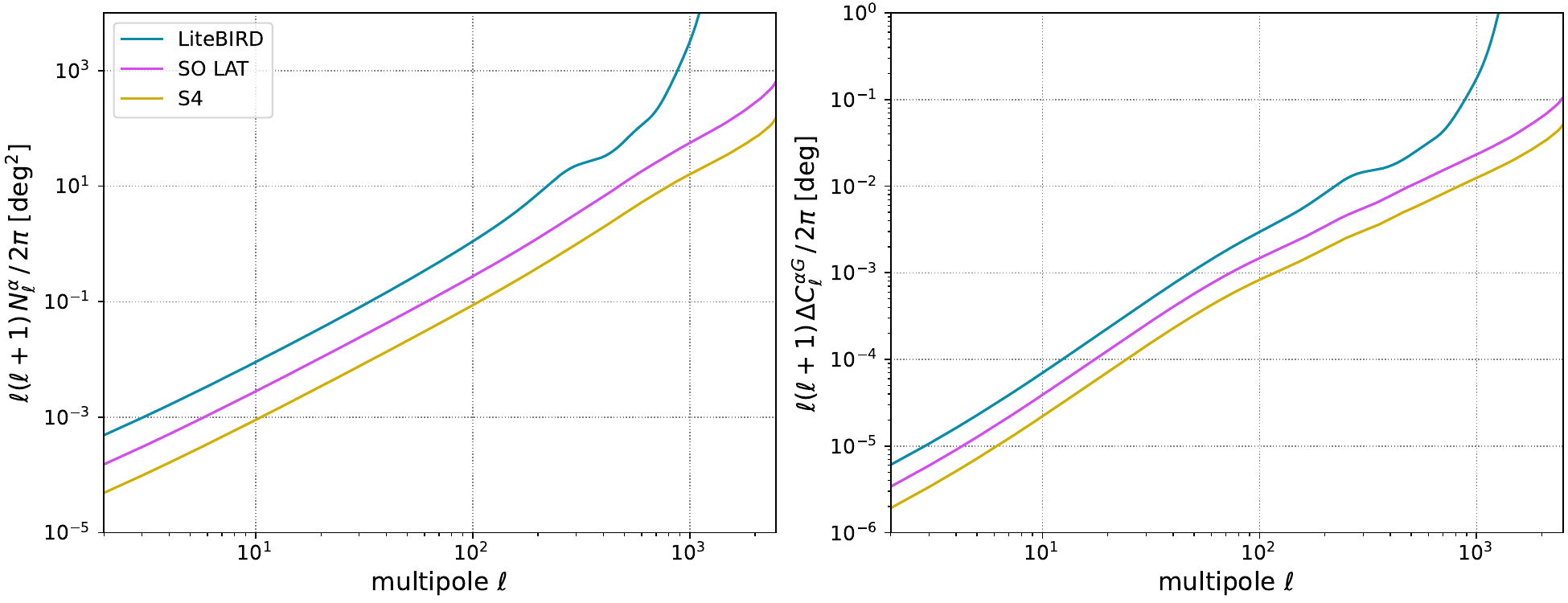}
\caption{(\emph{Left}): noise curves for the birefringence auto-correlation $C_\ell^{\alpha\alpha}$, computed through the harmonic estimator defined in \citep{Gluscevic:2009mm,Zagatti:2024jxm}, with the instrument specifications of table~\ref{tab:noise}, for the three related CMB experiments. (\emph{Right}): noise dominated variance of the birefringence-galaxy cross-correlation ($C_\ell^{\alpha G}$). The result refers to eq.~\eqref{eq:deltacross}, assuming $\alpha=0$. For galaxies we refer to the instrument specifications in the text below eq.~\eqref{eq:deltacross}.}
\label{fig:noise}
\end{figure}
\begin{alignat}{2}
    \notag\Delta_\ell(z,\,k) &= \Delta^{\textrm{Den}}_\ell(z,\,k,\,b_g,\,S_D)\quad&\textrm{(Density term)}\\
    &+ \notag\Delta^{\textrm{Len}}_\ell(z,\,k,\,S_{\Phi+\Psi})\quad&\textrm{(Lensing term)}\\
    \label{eq:Wg}&+ \left(\Delta^{\textrm{D1}}_\ell+\Delta^{\textrm{D2}}_\ell\right)(z,\,k,\,H,\,S_\Theta)\quad&\textrm{(Doppler term)}\\
    &+ \notag\Delta^{\textrm{Red}}_\ell(z,\,k,\,H,\,S_\Theta)\quad&\textrm{(RSD term)}\\
    &+ \notag\left(\Delta^{\textrm{G1}}_\ell+\ldots+\Delta^{\textrm{G5}}_\ell\right)(z,\,k,\,H,\,H',\,S_\Phi,\,S_\Psi,\,S_{\Phi+\Psi},\,S_\Theta)\;.\quad&\textrm{(Gravity term)}
\end{alignat}
Here, $b_g$ is the galaxy bias, $S_D$ a source function related to the growth of structure and $S_\Theta$ the velocity source function, while $S_\Phi$, $S_\Psi$, $S_{\Phi+\Psi}$ are related to the Bardeen potentials $\Phi$ and $\Psi$. Finally, $H$ is the Hubble parameter, and $H'$ its derivative with respect to conformal time. Eq.~\eqref{eq:Wg} accounts for the usual density term, lensing and Doppler effects. All redshift space distortions (RSD) are embedded in the fourth line, whereas the last term includes gravity corrections, among which we have the integrated Sachs-Wolfe effect. It now becomes evident how both eq.~\eqref{eq:alphakernel} and eq.~\eqref{eq:galkernel} are driven by the gravitational potentials, and thus the metric scalar perturbations, underscoring the indisputable relation between the two effects, if, how we show in the following, the temporal dependence allows the two kernels to peak each other out.
\begin{figure}[t]
\centering
\includegraphics[width=0.95\hsize]{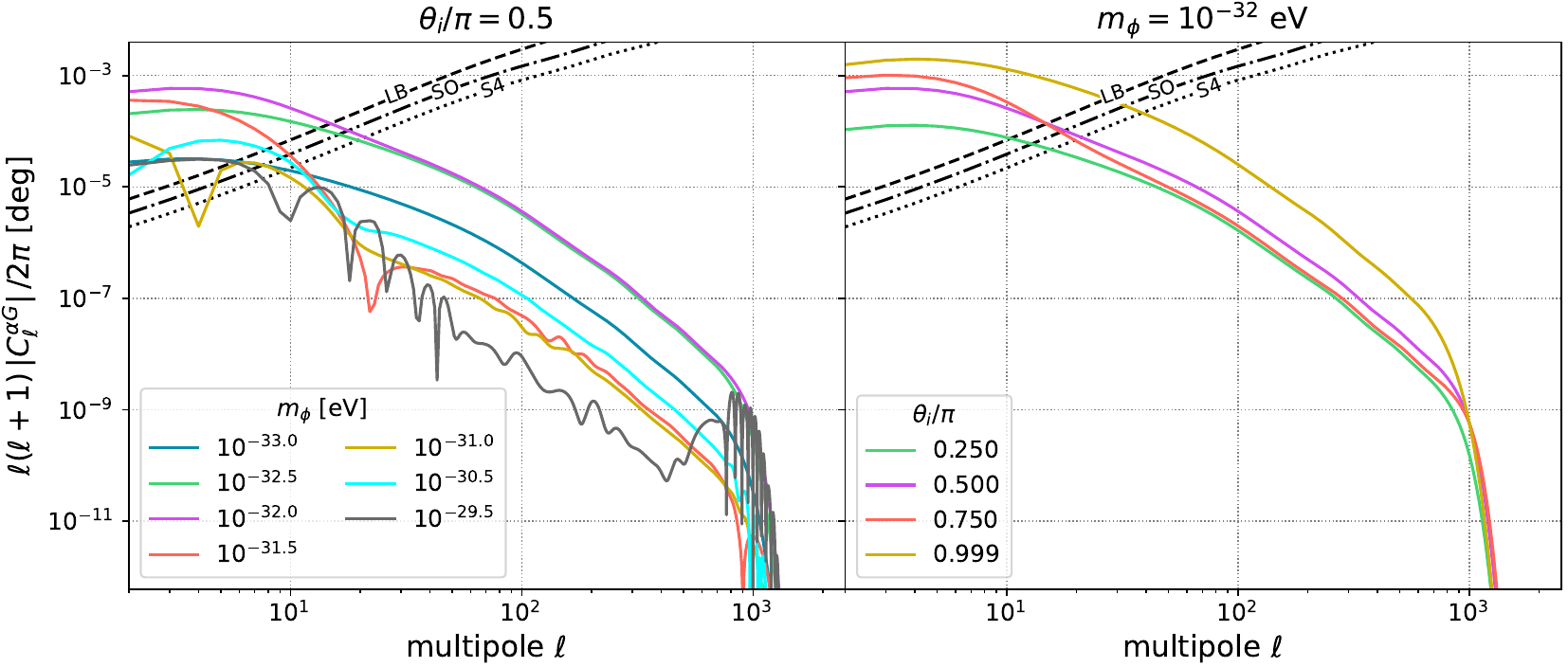}
\caption{Cross-correlation angular power spectrum between anisotropic birefringence and galaxy number counts for multiple values of the ALP mass and fixed initial misalignment angle $\theta_i=\pi/2$ (\emph{Left}), and multiple values of the initial misalignment angle and fixed mass $m_\phi=10^{-32}\,\si{\electronvolt}$ (\emph{Right}). The black lines refer to the noise levels of eq.~\eqref{eq:deltacross} for \emph{Euclid} as a galaxy survey and the CMB experiments under consideration: CMB-S4 (dashed), Simons Observatory (dot-dashed) and LiteBIRD (dotted). All spectra are computed with an axion-photon coupling $g_{\phi\gamma}=\SI{2e-14}{\giga\electronvolt^{-1}}$, and scale linearly with increasing coupling values.}
\label{fig:cells}
\end{figure}

Similarly, from the kernels in eq.~\eqref{eq:alphakernel} and eq.~\eqref{eq:galkernel} we obtain the birefringence and galaxy auto angular power spectra, respectively,
\begin{align}
    \label{eq:alphaalpha}C^{\alpha\alpha}_\ell &= 4\pi\int\frac{{\rm d}k}{k}\;\mathcal{P}_\mathcal{R}(k)\left(\Delta_\ell^\alpha(k)\right)^2\\
    \label{eq:GG}C_\ell^{G_i G_j} &= 4\pi\int \frac{{\rm d}k}{k}\;\mathcal{P}_\mathcal{R}(k)\Delta_\ell^{G_i}(k)\Delta_\ell^{G_j}(k) \;,
\end{align}
where the latter accounts for correlations between the $i$-th and $j$-th redshift bin.

The cross-correlation of eq.~\eqref{eq:alphaG}, the two auto-correlations in eqs.~\eqref{eq:alphaalpha}-\eqref{eq:GG} and the underlying axion model (described in previous sections) have been implemented in a properly modified version of the Boltzmann code {\tt CLASS} \citep{Lesgourgues:2011re,Blas:2011rf}, following the prescription of {\tt CLASSgal} \citep{DiDio:2013bqa}, for galaxies, and those of ref.~\citep{Greco:2022xwj} for birefringence.

Finally, let us define the variance of the correlations $(\Delta C_\ell^{XY})^2\equiv \mathrm{Cov}(C_\ell^X,\;C_\ell^Y)$, which will be useful in our analysis,
\begin{align}
        \label{eq:deltacross}\left(\Delta C_\ell^{\alpha G_{ij}}\right)^2&=\frac{1}{(2\ell+1)f_{\rm sky}^{\alpha G}}\left[C_\ell^{\alpha G_i}C_\ell^{\alpha G_j}+\left(C_\ell^{\alpha\alpha}+N_\ell^\alpha\right)\left(C_\ell^{G_iG_j}+N^G_{ij}\right)\right] \;,\\
        \label{eq:deltaauto}\left(\Delta C_\ell^{\alpha\alpha}\right)^2&=\frac{2}{(2\ell+1)f_{\rm sky}^\alpha}\left(C_\ell^{\alpha\alpha}+N_\ell^\alpha\right)^2 \;.
\end{align}
Here, $f_{\rm sky}^{\alpha G}$ is the effective sky fraction observed by the combined analysis of the CMB telescope and the galaxy survey of interest (whose sky fraction read $f_{\rm sky}^\alpha$ and $f_{\rm sky}^G$, respectively), taken in the following as the lowest of the two. For the former refer to table~\ref{tab:noise}, whilst for the latter we take the forecast DR3 fraction ($36\%$). $N_\ell^\alpha$ and $N^G_{ij}$ are the related noise-level, respectively. The multipole dependence of $N_\ell^\alpha$ stems from how the angular resolution of the underlying CMB experiment varies in harmonic space. In contrast, galaxy surveys have achieved an extremely high angular resolution, such that their beam function can be considered equal to unity over the entire range of scales under consideration. The galaxy shot noise is inversely proportional to the average number of measured galaxies per angular volume projected for the target survey ($N^G_{ij}=n_{\rm bin}/\bar N_g\,\delta_{ij}$, with $\bar N_g^{\rm Euclid}=30\text{ arcmin}^{-2}$ \citep{Euclid:2019clj} and $n_{\rm bin}$ the number of selected redshift bins). Meanwhile, the term $N_\ell^\alpha$ is computed following the approach of refs.~\citep{Gluscevic:2009mm,Zagatti:2024jxm}, relying on an harmonic estimator for birefringence constructed onto the CMB E and B mode cross-correlation. The left panel of fig.~\ref{fig:noise} shows the related noise curves for LiteBIRD, the Simons Observatorty LAT and CMB-S4, whose sensitivity on polarization $\sigma_P$ and angular resolution $\theta_{\rm fwhm}$ are reported in table~\ref{tab:noise}. Alongside, in the right panel we illustrate the noise-dominated variance on the cross-correlation, without tomography, i.e. using a single redshift bin that goes from $z=0.001$ up to $z=2.5$. This is computed from eq.~\eqref{eq:deltacross}, assuming $C_\ell^{\alpha\alpha}=C_\ell^{\alpha G} = 0$.\footnote{This is a suitable approximation as long as the axion-photon coupling remains fairly low, resulting in $C_\ell^{\alpha G}\ll C_\ell^{\alpha \alpha} \ll N_\ell^\alpha$.} The results outlined in the following section are also obtained through the non-tomographic approach, hence, we will be dropping the superscript referred to the $i$-th redshift bin from now on. Refer to appendix~\ref{app:B} for details on the impact of tomography and the dominant redshift-range of the underlying cross-correlation. 

\section{Results}
\label{sec:res}
The aim of this work is to present the cross-correlation between axion induced cosmic birefringence and galaxy counts as a novel tool to constraint axion-like physics. Previous works have extensively studied the cross-correlation of birefringence and CMB to test early dark energy models as a possible solution to the Hubble tension \citep{Capparelli:2019rtn, Poulin:2018dzj} and explore the impact of the axion mass on the amplitude of the rotation effect \citep{Greco:2022xwj,Greco:2024oie}. As discussed in section \ref{subsec:bkg}, masses higher than $\sim10^{-28}\,\si{\electronvolt}$ induce oscillations in the field evolution prior to recombination, leading to a fainter birefringence signature due to the smaller difference between the field's present-day value and its value at the time of last scattering, see eq.~\eqref{eq:alpha_inst}, and to an even fainter signal for the birefringence from reionization. Moreover, lower masses have proven effective in reproducing the angle constrained by CMB data \citep{Minami:2020odp,Diego-Palazuelos:2022dsq,Eskilt:2022wav,Eskilt:2022cff}, as shown in refs.~\citep{Greco:2022xwj,Galaverni:2023zhv,Greco:2024oie, Fujita:2020ecn}. For these reasons and because of the numerical complications of dealing with higher masses\footnote{When $m_\phi>10^{-28}\,\si{\electronvolt}$, oscillations not only begin before the epoch of recombination but also accelerate as they progress towards the later stages of the Universe. Consequently, {\tt CLASS} encounters numerical challenges when employing very small integration time-steps, leading to significantly longer execution times required to achieve the necessary resolution for the cross-correlation under investigation.}, we concentrate our analysis on $10^{-33}\,\si{\electronvolt}\le m_\phi\le10^{-28}\,\si{\electronvolt}$. Additionally, galaxies serve as low-redshift tracers, implying that we anticipate larger contributions to the cross-correlation for fields whose evolution commences at later epochs, associated with lower masses. Beyond mass considerations, we explore variations in the initial misalignment angle $\theta_i$, establishing the "in vacuo" initial condition post-symmetry breakdown, and in the axion-photon coupling $g_{\phi\gamma}$, which determines the amplitude of the cross-correlation signal.

The level of coupling between axions and photons has been constrained by haloscopes, helioscopes, colliders and astrophysical searches (see {\tt AxionLimits} \citep{axionlimits} for all data and references) down to $m_\phi\sim10^{-12}\,\si{\electronvolt}$. Ref.~\citep{Galaverni:2023zhv} presents a novel constraint down to $m_\phi\sim10^{-24}\,\si{\electronvolt}$, derived from isotropic birefringence\footnote{Ref. \citep{Gan:2023swl} derives a bound, in a similar mass range, based on cosmic birefringence around supermassive black holes.}, while ref.~\citep{Choi:2021aze} within a model of electroweak axion dark energy, accounting for the non-vanishing birefringence angle of ref.~\citep{Minami:2020odp}, at masses around $10^{-33}\,\si{\electronvolt}$. Similarly, ref.~\citep{Greco:2024oie} constraints the axion-parameters for a mass range $m_\phi\in\left(10^{-33},\;10^{-26}\right)\,\si{\electronvolt}$, exploiting the isotropic birefringence angle of ref. \citep{Diego-Palazuelos:2022dsq}. Nonetheless, the targeted masses of our analysis remain poorly constrained due to their ultralight nature, far below the sensitivities of direct search experiments, and cosmology offers an excellent avenue for probing this region of the parameter space. On the account of these considerations, we adopt $g_{\phi\gamma}=10^{-12}\,\si{\giga\electronvolt^{-1}}$ as an upper bound of our parameter space \citep{axionlimits,Reynes:2021bpe,Reynolds:2019uqt}. In summary, our focus lies within the following parameter range: $10^{-33}\,\si{\electronvolt}\le m_\phi\le10^{-28}\,\si{\electronvolt}$, $-\pi<\theta_i\le\pi$, $g_{\phi\gamma}\le10^{-12}\,\si{\giga\electronvolt^{-1}}$ (refer to section \ref{subsec:potential} for a discussion on the range of variation of the initial misalignment angle $\theta_i$).

\begin{figure}[t]
\centering
\includegraphics[width=0.95\hsize]{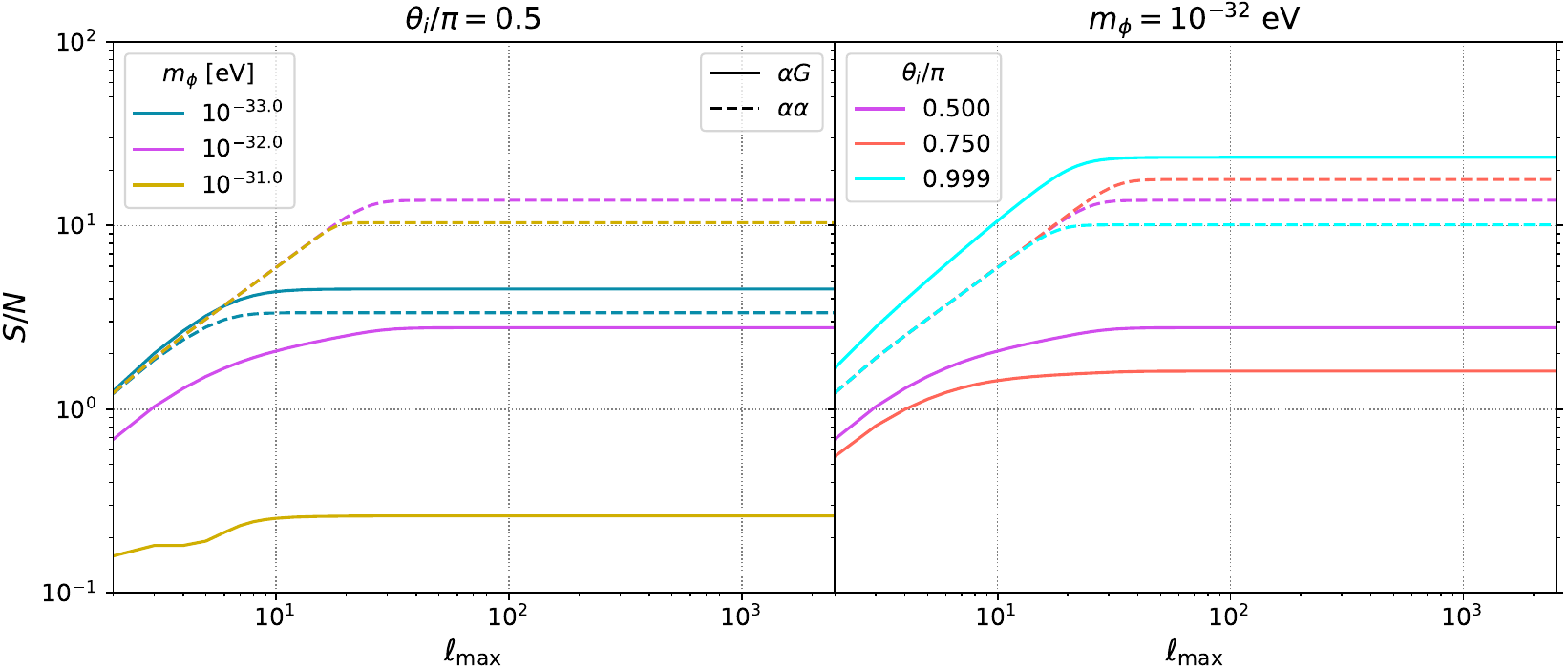}
\caption{Signal-to-noise ratios of the birefringence-galaxy cross-correlation (solid) and birefringence auto-correlation (dashed) as a function of the maximum multipole $\ell_{\rm max}$. The result is shown separately for multiple points in the $m_\phi\,$-$\,\theta_i$ parameter space at fixed misaligment angle and mass in the left and right panel, respectively (the axion-photon coupling is set to $\SI{2e-14}{\giga\electronvolt^{-1}}$).}
\label{fig:snr_ell_notomo}
\end{figure}

We have computed the cross-correlation signal with our modified version of {\tt CLASS}, following the prescriptions of section \ref{sec:cross} and accounting for an EDE model of the axion potential with $n=3$, as discussed in section \ref{subsec:potential}. For the following analysis, the angular power spectra are calculated in a non-tomographic framework, specifically for a single redshift bin up to $z=2.5$. Nonetheless, we discuss the impact of tomography in appendix \ref{app:B}, exploiting the ten redshift bins of \emph{Euclid}~\citep{Euclid:2019clj}.\footnote{We expect the signal to grow as the redshift approaches the peak of the non-integrated galaxy kernel in eq.~\eqref{eq:galkernel}. However, this trend is influenced by a complex interplay between the ALP-parameters and the latter. For certain points in the parameter space, the cross-correlation peaks at redshifts exceeding 2.5, which may be explored by surveys other than the one currently under consideration, namely \emph{Euclid}.}
Fig.~\ref{fig:cells} illustrates the cross-correlation alongside with the corresponding variance from eq.~\eqref{eq:deltacross} for the CMB experiments considered in our study (S4, SO and LB). All the spectra in that figure have been computed with the axion-photon coupling fixed to $g_{\phi\gamma}=\SI{2e-14}{\giga\electronvolt^{-1}}$.\footnote{By looking at eq.~\eqref{eq:alphakernel}, it is evident that the cross-correlation increases linearly with the axion-photon coupling. Hence, as the latter grows, we anticipate the signal to stand out from the related variance, even at higher multipoles.} 
In the left panel, we illustrate the correlation's dependency on the field's mass, with the initial misalignment angle fixed at $\theta_i=\pi/2$. The signal tends to peak at large angular scales, with most of the signal-dominated information coming from $\ell<100$. Subsequently, the signal starts to decrease, transitioning to an oscillatory pattern around zero for larger masses. This oscillation occurs at lower multipoles as the mass increases, stemming from a complicated interplay between the extended duration of oscillations and the galaxy tracer transfer function. 
The signal's amplitude peaks around $m_\phi\sim10^{-32}\,\si{\electronvolt}$, decreasing for masses smaller or larger than this value. The decrease is however less pronounced for smaller than for larger masses. We anticipate a stronger correlation for axion-like fields whose transfer function aligns with the peak of the galaxy transfer function. Given that galaxies serve as low-redshift tracers, fields initiating their evolution at later epochs are preferred (i.e., smaller masses). In the right panel of fig.~\ref{fig:cells}, we show how the cross-correlation varies with respect to the initial misalignment angle, at fixed mass $m_\phi=10^{-32}\,\si{\electronvolt}$. A larger angle corresponds to a greater amplitude of the correlation. The initial misalignment angle not only determines the initial positioning of the field within its potential (see eq.~\eqref{eq:potential}), thereby altering its oscillatory behavior, but also establishes the initial value of the field itself and subsequently influences the magnitude of its evolution. This effect is particularly pronounced for small masses, where oscillations have yet to commence today, and the field's value remains close to its initial condition.

Let us underline that the parameters discussed in eqs.~\eqref{eq:alpha_inst} and \eqref{eq:alpha} play a crucial role in predicting both the isotropic and anisotropic components of the birefringence angle. To this end, it is useful to explore the region of parameter space able to reproduce the isotropic angle, as constrained by available polarization data \citep{Minami:2020odp,Eskilt:2022cff,Eskilt:2022wav,Diego-Palazuelos:2022dsq}. An extensive analysis has been performed by some of us in ref.~\citep{Greco:2024oie} for quadratic potentials, identifying a set of best fit parameters yielding the observed rotation angle ($\alpha_0=0.3^\circ\pm0.11^\circ$ \citep{Diego-Palazuelos:2022dsq}). Additionally, figure (5a) of ref.~\citep{Greco:2024oie} shows the constrained 2-$\sigma$ region in the parameter space $m_\phi$-$g_{\phi\gamma}$ for the same EDE model under consideration in this work. However, the initial misalignment angle is fixed in that figure, as the isotropic component of birefringence is degenerate in the product $g_{\phi\gamma}\,\theta_i$. This does not hold for the cross-correlation presented here, offering as an excellent tool to break this degeneracy but, at the same time, making a direct comparison with the results of~\citep{Greco:2024oie} not completely straightforward. For this reason, in the following we explore the axion-like parameter space mostly focusing on the detectability of the cross-correlation alone. We will anyway provide a more detailed comparison with ~\citep{Greco:2024oie} below.

\subsection{Exploring the ALP-parameter space with the signal-to-noise ratio}
\label{subsec:snr}
\begin{figure}[t]
\centering
\includegraphics[width=0.95\hsize]{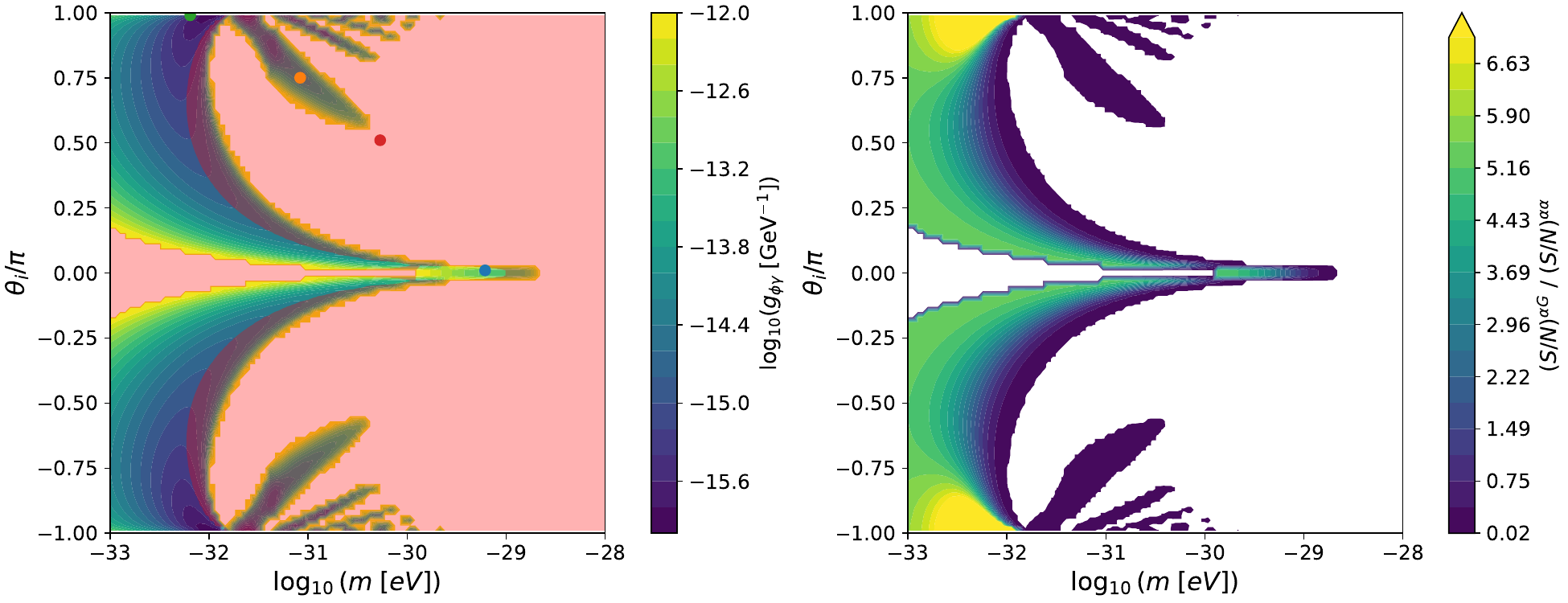}
\caption{The signal-to-noise ratios presented here are computed with the instrument specifications of CMB-S4 (see table~\ref{tab:noise}) and those of \emph{Euclid} presented in section~\ref{sec:cross}. We set the minimum observable multipole to $\ell_{\rm min}=10$. (\emph{Left}): axion-photon coupling $g_{\phi\gamma}$ needed to achieve unity in the $(S/N)$ of the cross-correlation. The four dots correspond to particular points of the parameter space, for which we show the $(S/N)$-coupling relation in fig.~\ref{fig:4cases}. (\emph{Right}): ratio between the $(S/N)$ of the cross-correlation and the auto-correlation for the corresponding coupling values needed to achieve unity in $(S/N)^{\alpha G}$, for each point of the parameter space in the left panel. The red-shaded region identifies where this ratio is smaller than one, whilst missing points around the contour refer to where $(S/N)^{\alpha G}<1$ even for the largest coupling considered in our analysis ($g_{\phi\gamma}=10^{-12}\,\si{\giga\electronvolt^{-1}}$). Let us stress how the constrained region is compatible, to a vast extent, with that able to reproduce the most recent measurements of the isotropic birefringence angle \citep{Greco:2024oie}.}
\label{fig:bound_s4}
\end{figure}
We aim to determine the potential detectability of the signal by combining upcoming data releases from the \emph{Euclid} survey \citep{Euclid:2019clj} with future and forthcoming polarization data from LiteBIRD \citep{LiteBIRD:2022cnt}, SO \citep{SimonsObservatory:2018koc} and S4 \citep{CMB-S4:2020lpa}. To this purpose, we explore variations of the signal-to-noise ratio in the entire parameter space ($m_\phi$, $\theta_i$, $g_{\phi\gamma}$). We compute the $(S/N)$ both for the cross-correlation $C_\ell^{\alpha G}$ and auto-correlation $C_\ell^{\alpha\alpha}$. Our goal is to identify regions where the cross-correlation surpasses a threshold of one, indicating potential detectability in experiments, and where it also exceeds the auto-correlation. This suggests that the cross-correlation offers potentially richer information compared to the auto-correlation alone. The corresponding signal-to-noise ratios read
\begin{figure}[t]
\centering
\includegraphics[width=0.95\hsize]{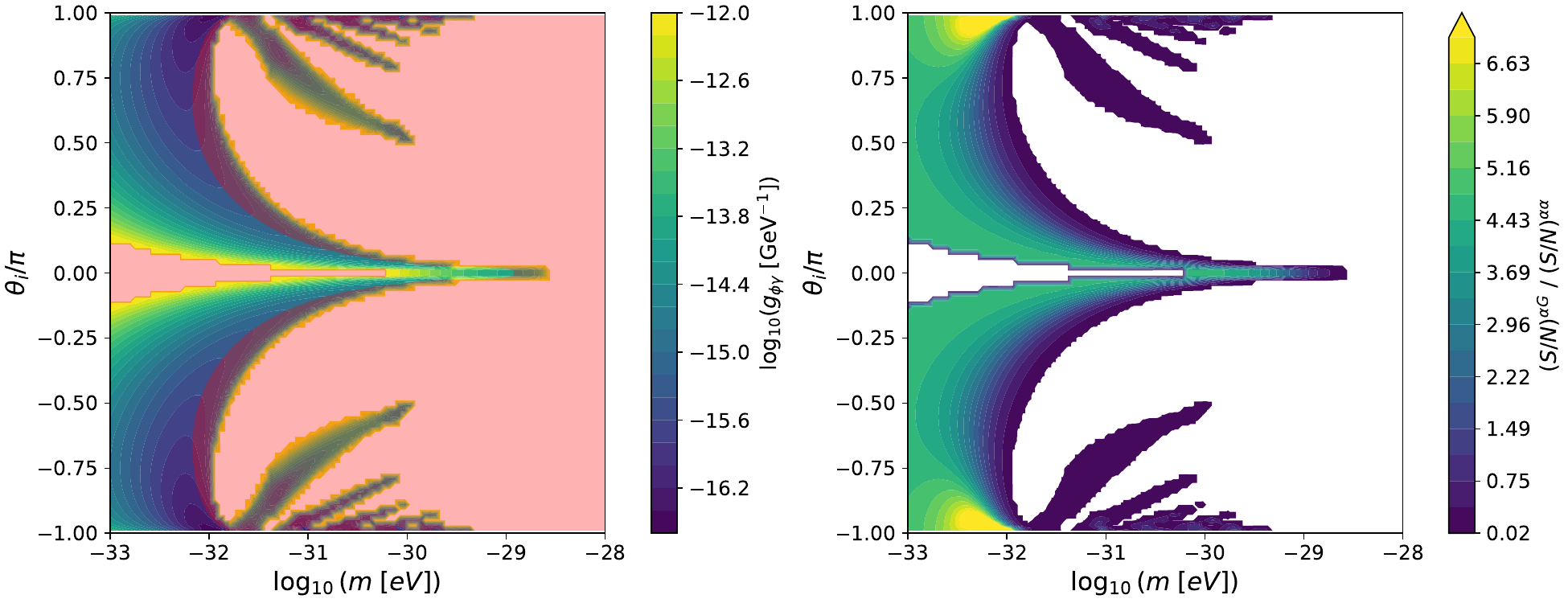}
\caption{Same as fig.~\ref{fig:bound_s4}, with the minimum multipole of the signal-to-noise ratios in eqs.~\eqref{eq:S/Ncross}~and~\eqref{eq:S/Nauto} brought down to 5.}
\label{fig:bound_ellmin10}
\end{figure}
\begin{align}   
    \label{eq:S/Ncross}(S/N)^{\alpha G} &= \sqrt{\sum_{\ell=\ell_{\rm min}}^{\ell_{\rm max}}\left(\frac{C_\ell^{\alpha G}}{\Delta C_\ell^{\alpha G}}\right)^2} \;,\\
    \label{eq:S/Nauto}(S/N)^{\alpha\alpha} &= \sqrt{\sum_{\ell=\ell_{\rm min}}^{\ell_{\rm max}}\left(\frac{C_\ell^{\alpha\alpha}}{\Delta C_\ell^{\alpha\alpha}}\right)^2} \;,
\end{align}
$(S/N)^{\alpha G}$ is computed in the non-tomographic scenario, focusing on a single redshift bin up to $z=2.5$. It's worth mentioning briefly that we anticipate a slight improvement in the results when applying tomography and refer to appendix \ref{app:B} for further discussion. In eqs.~\eqref{eq:S/Ncross}~and~\eqref{eq:S/Nauto}, $\ell_{\rm max}$ is the maximum multipole up to which the targeted experiments are sensitive to this novel cross-correlation. Although we fix it to the highest multipole computed by {\tt CLASS}, let us notice that the $(S/N)$ saturates at smaller multipoles. Examining fig.~\ref{fig:cells} we observe how the cross-correlation peaks at large angular scales (small $\ell$'s), with its variance notably exceeding the signal around $\ell=30-40$.\footnote{The interplay between the signal and its variance is strongly dependent on the axion-photon coupling. As the latter increases, the variance surpasses the signal at higher multipoles} Fig.~\ref{fig:snr_ell_notomo} depicts the behavior of eqs.~\eqref{eq:S/Ncross}~and~\eqref{eq:S/Nauto} with respect to $\ell_{\rm max}$ for multipole points in the $m_\phi\,$-$\,\theta_i$ parameter space and $g_{\phi\gamma}=\SI{2e-14}{\giga\electronvolt^{-1}}$.\footnote{Here, we set $\ell_{\rm min}=2$ to show the impact of each multipole on the signal-to-noise ratio. However, in the subsequent analysis, we will adopt a higher minimum multipole in accordance with \emph{Euclid}'s guidelines \citep{Euclid:2019clj}.} The multipole necessary for the $(S/N)$ to saturate significantly depends on the ALP-parameters. However, it is reasonable to presume that all of the information is enclosed before $\ell=100$.

Wide galaxy surveys such as \emph{Euclid} are not anticipated to probe the largest scales due to their expected sky coverage. Following \citep{Euclid:2019clj}, we consider $\ell_{\rm min}=10$ as our baseline, while also considering a more optimistic scenario where scales down to $\ell_{\rm min}=5$ can be observed. Fig.~\ref{fig:snr_ell_notomo} also demonstrates that the initial multipoles do not significantly impact the $(S/N)$, as most of the contribution is added in for $\ell>10$ especially for lower masses and high initial misalignment angle.

\begin{figure}[t]
\centering
\includegraphics[width=0.95\hsize]{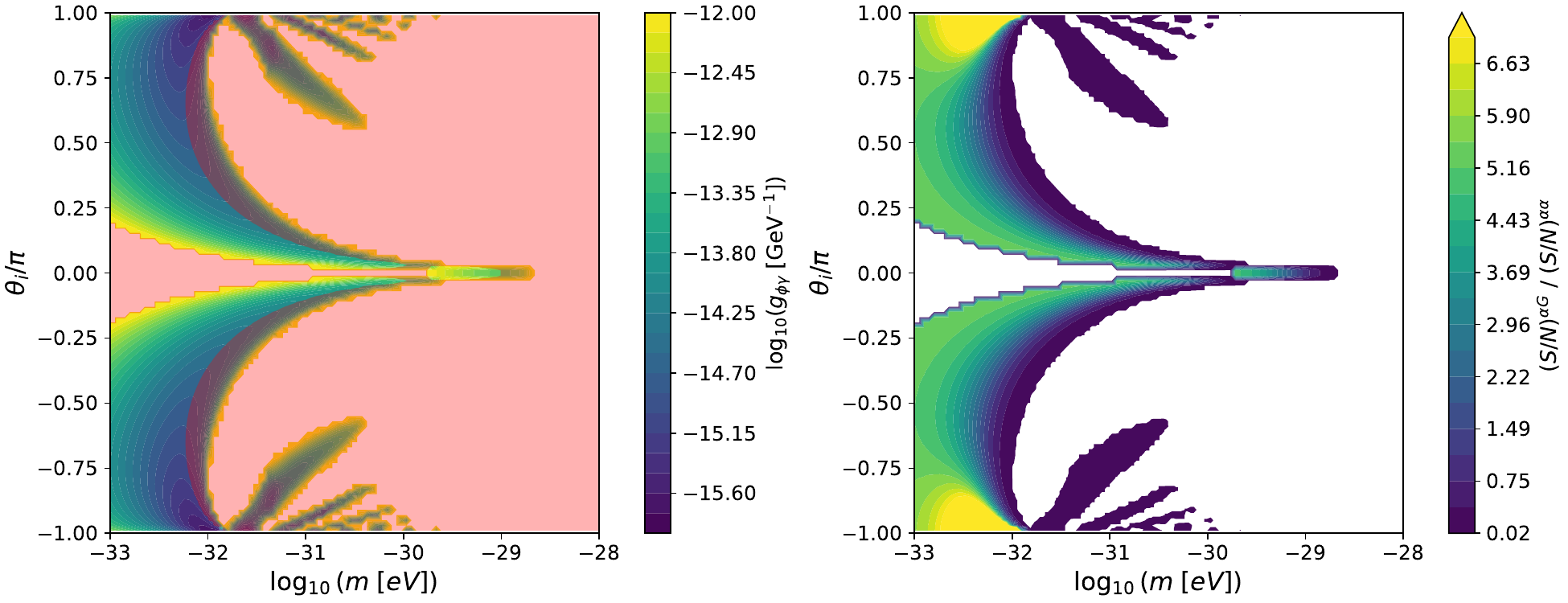}
\caption{Same as fig.~\ref{fig:bound_s4}, with the instrument specifications of the Simons Observatory (see table~\ref{tab:noise}).}
\label{fig:bound_so}
\end{figure}

We ran our modified version of {\tt CLASS} on a $100\times100$ grid in the $\theta_i$-$m_\phi$ parameter space\footnote{In doing so, we ensured that the spectator field approximation always holds, by checking that the axion energy density is negligible with respect to standard $\Lambda$CDM components. Further insights can be found in appendix~\ref{app:C}.} and found the necessary value of the axion-photon coupling $g_{\phi\gamma}$ to get $(S/N)^{\alpha G}>1$. The latter has a strong impact on the signal-to-noise ratio, as it appears as a multiplicative factor in the cross-correlation (see eq.~\eqref{eq:alphaG}) and as its square in the auto-correlation (see eq.~\eqref{eq:alphaalpha}), stemming from the angular coefficients of eq.~\eqref{eq:alm}. As long as the birefringence noise term dominates over the related auto-correlation in eqs.~\eqref{eq:S/Ncross}-\eqref{eq:S/Nauto}, $(S/N)^{\alpha G}\propto g_{\phi\gamma}$ and $(S/N)^{\alpha\alpha}\propto g^2_{\phi\gamma}$. However, this relationship no longer holds as the coupling increases. Eventually, when the auto-correlation significantly surpasses the noise, both signal-to-noise ratios will saturate to a $g_{\phi\gamma}$-independent value, since their numerator and denominator scale with the same power. This asymptotic behavior can be summarized as follows:
\begin{align}   
    \label{eq:cross_sat}(S/N)^{\alpha G} &\longrightarrow\sqrt{\sum_{\ell=\ell{_{\rm min}}}^{\ell_{\rm max}}\frac{\left(\tilde C_\ell^{\alpha G}\right)^2}{\tilde C_\ell^{\alpha \alpha}\left(C_\ell^{GG}+N^G\right)}(2\ell+1)f_{\rm sky}^{\alpha G}} \;,\\
    \label{eq:auto_sat}(S/N)^{\alpha\alpha} &\longrightarrow \sqrt{\sum_{\ell=\ell_{\rm min}}^{\ell_{\rm max}}\frac{2\ell+1}{2}f_{\rm sky}^\alpha} \;,
\end{align}
where $\tilde C_\ell^{\alpha G}=C_\ell^{\alpha G}/g_{\phi\gamma}$ and $\tilde C_\ell^{\alpha\alpha}=C_\ell^{\alpha\alpha}/g^2_{\phi\gamma}$ are the coupling-independent correlations. It's notable that this asymptotic value of the signal-to-noise ratio is solely dependent on the underlying ALP-parameters ($m_\phi$ and $\theta_i$). Moreover, the cross-correlation is always disfavoured with respect to the auto-correlation, when saturated. Predicting beforehand when this occurs in terms of coupling is challenging. It may even happen after the coupling exceeds $g_{\phi\gamma}=10^{-12}\,\si{\giga\electronvolt^{-1}}$ (the upper limit set for our analysis, as discussed earlier in this section), potentially leaving a portion of the parameter space where a coupling value, within current constraints, yields a signal-to-noise ratio exceeding both unity and that of the auto-correlation.

The left panel of fig.~\ref{fig:bound_s4} illustrates the required coupling value to achieve a $(S/N)$ of order unity for the cross-correlation, allowing values smaller than $g_{\phi\gamma}=10^{-12}\,\si{\giga\electronvolt^{-1}}$ only. The signal-to-noise is computed with the instrument specifications of CMB-S4, as of table~\ref{tab:noise}, and the \emph{Euclid} prescriptions of section~\ref{sec:cross}. Following \citep{Euclid:2019clj}, we set the minimum observable multipole to 10, as wide galaxy surveys are unlikely to observe the largest scales. The cross-correlation offers a measurable probe of birefringence, for masses $m_\phi\lesssim10^{-32}\,\si{\electronvolt}$ and\footnote{These masses fall within the range where we anticipate a significant contribution to birefringence from photons emitted during the reionization period \citep{Greco:2022xwj,Nakatsuka:2022epj}. Therefore, a future data analysis of this innovative cross-correlation could not only constrain the ALP-parameter space but also provide valuable insights into the origin of birefringence.} initial misalignment angles $|\theta_i|\gtrsim\pi/4$. In this range, a coupling value exists within current bounds, resulting in a signal-to-noise ratio exceeding unity. The region of interest extends to slightly lower values of $|\theta_i|$ as the mass increases up to $10^{-30.5}\,\si{\electronvolt}$ and to a very small window around $m_\phi=10^{-29}\,\si{\electronvolt}$ when the initial misalignment angle is smaller than $\pi/100$. Let us stress that larger coupling values would yield larger signal-to-noise ratios, rendering regions where the required coupling value is minimal more promising in terms of detectability with forthcoming surveys. To this end, the most promising result is obtained for $m_\phi\sim10^{-32}\,\si{\electronvolt}$ and $|\theta_i|\sim\pi$, whereas, for smaller masses, the effect is less prominent as the amplitude of the underlying cross-correlation lowers (see fig.~\ref{fig:cells}). Additionally, in the right panel of fig.~\ref{fig:bound_s4}, we present the ratio, in signal-to-noise, between the cross- and auto-correlation, corresponding to the axion-photon coupling that results in $(S/N)^{\alpha G}=1$ for each point of the examined parameter space. Smaller masses and larger initial misalignment angles lead to higher ratios, indicating that the cross-correlation provides more information than the auto-correlation in this domain. In particular, we highlight this behavior in the left panel, by excluding the region of parameter space where this ratio is lower than unity (the red-shaded region). Also, in fig.~\ref{fig:bound_ellmin10} we assess the impact of decreasing the minimum multipole in the computation of the singal-to-noise-ratios in eqs.~\eqref{eq:S/Ncross}~and~\eqref{eq:S/Nauto}, adopting a more optimistic scenario where $\ell_{\rm min}=5$. The constrained region is slightly enhanced, and the axion-photon coupling value needed to achieve unity in the $(S/N)$ decreases. Nevertheless, as shown in fig.~\ref{fig:snr_ell_notomo}, most of the signal-dominated information comes from higher multipoles. Hence, we anticipate that a more thorough observational analysis able to probe scales below $\ell=10$ would not significantly improve our results.

\begin{figure}[t]
\centering
\includegraphics[width=0.95\hsize]{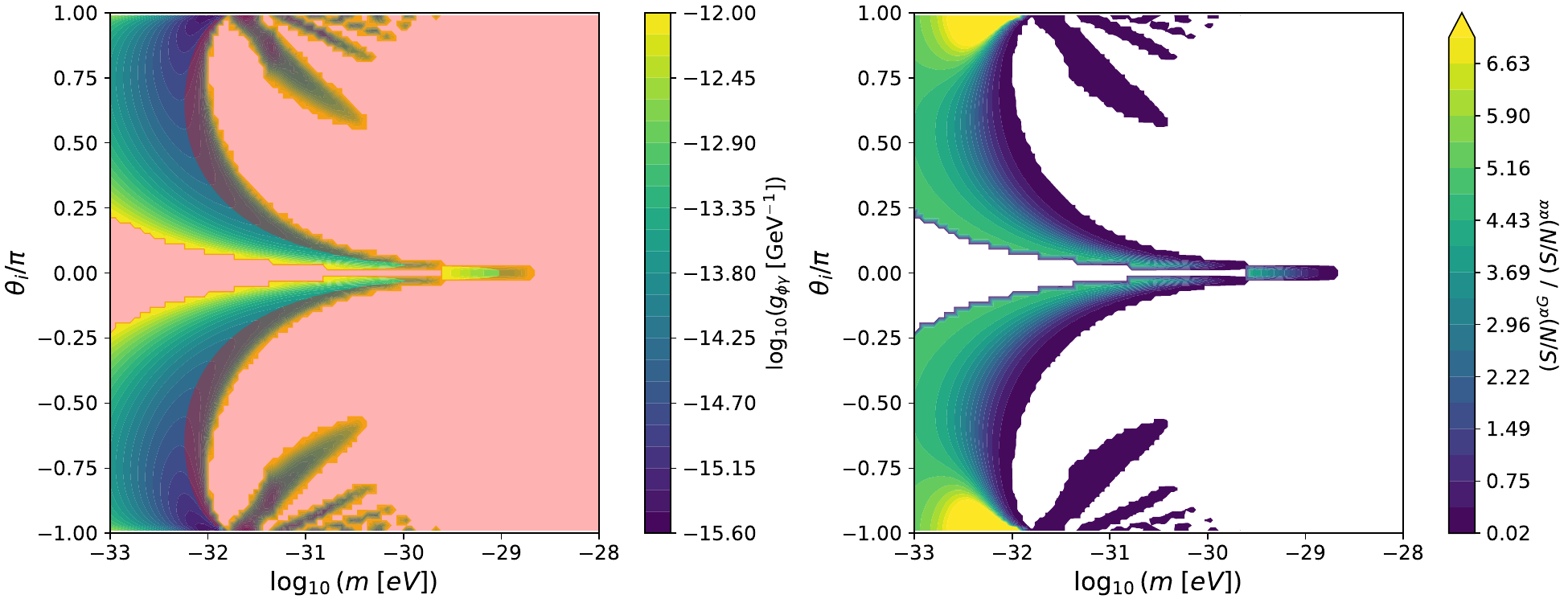}
\caption{Same as fig.~\ref{fig:bound_s4}, with the instrument specifications of the LiteBIRD (see table~\ref{tab:noise}).}
\label{fig:bound_litebird}
\end{figure}

\begin{figure}[t]
\centering
\includegraphics[width=0.95\hsize]{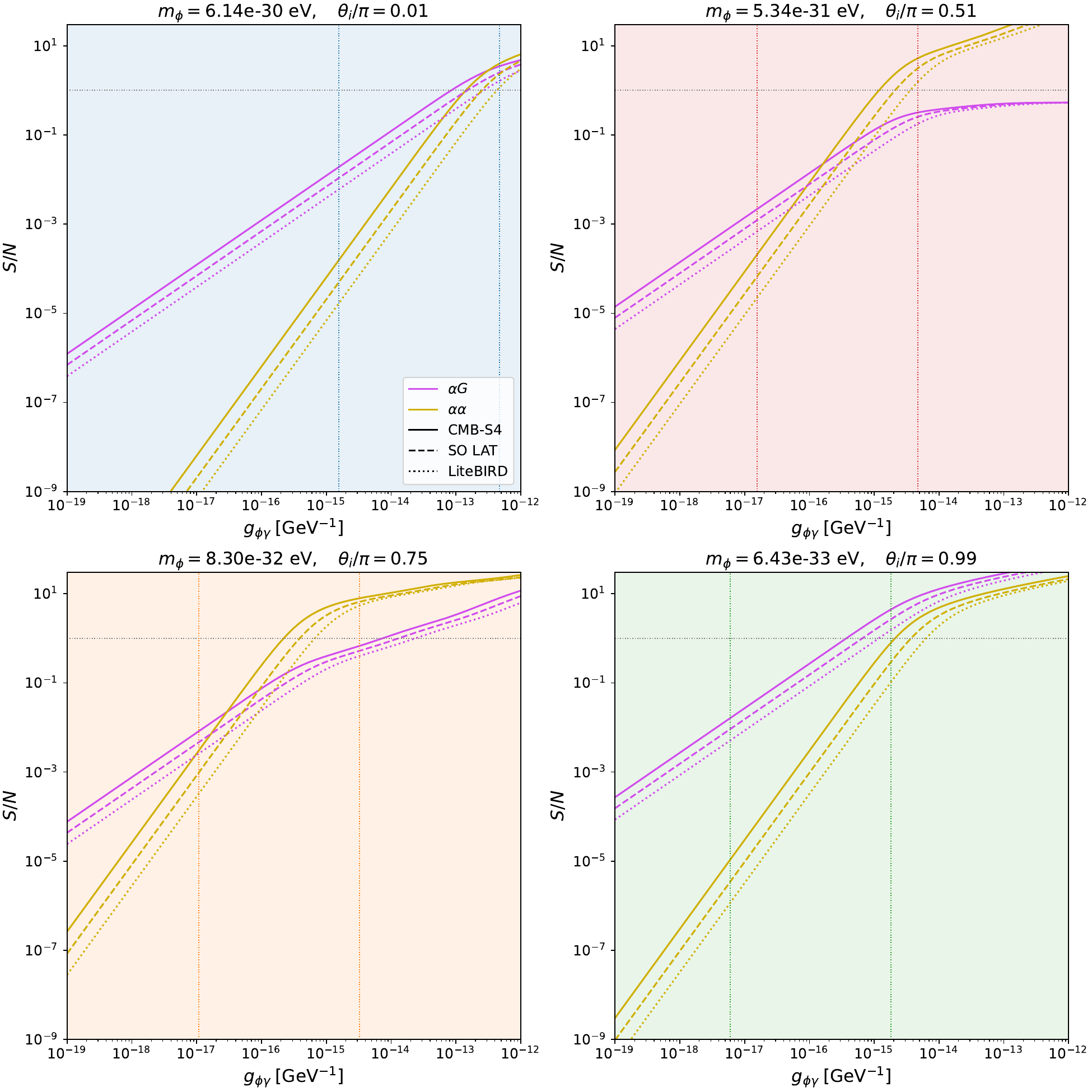}
\caption{Signal-to-noise ratio both for the cross- and auto-correlation as a function of the axion-photon coupling $g_{\phi\gamma}$, corresponding to four different points of the $m_\phi\,$-$\,\theta_i$ parameter space, highlighted with corresponding colors in fig.~\ref{fig:bound_s4}. For each panel, we show the result for all three CMB experiments under considerations: CMB-S4 (solid lines), the Simons Observatory (dashed lines) and LiteBIRD (dotted lines). (\emph{Top left}): the $(S/N)$ of the cross-correlation becomes larger than unity but gets surpassed by that of the auto-correlation afterwards. (\emph{Top right}): the $(S/N)$ of the cross-correlation never becomes larger than unity. (\emph{Bottom left}): the $(S/N)$ of the cross-correlation becomes larger than unity but gets surpassed by that of the auto-correlation beforehand. (\emph{Bottom right}): the $(S/N)$ of the cross-correlation becomes larger than unity and never gets surpassed by that of the auto-correlation. The two vertical lines in each panel correspond to the coupling value required to achieve the isotropic birefringence measurement of ref.~\citep{Diego-Palazuelos:2022dsq} and its first phase-shift (i.e., $\alpha_0=0.3^\circ$ for the left line and $\alpha_1=180.3^\circ$ for the right one).}
\label{fig:4cases}
\end{figure}

Figs. \ref{fig:bound_so} and \ref{fig:bound_litebird} illustrate the same result when using the Simons Observatory and LiteBIRD as target experiments, respectively. The primary distinction lies in the higher coupling value required to achieve unity in the signal-to-noise ratio across the entire parameter space. The excluded region is also enlarged as the polarization-sensitivity and angular resolution of the target experiment decrease. Additionally, it's worth noting that while ground-based experiments such as SO and CMB-S4 are capable of observing the sky down to $\ell\sim50$, information regarding large angular scales can still be derived from higher multipoles when constructing the birefringence maps  \citep{Bortolami:2022whx,Zagatti:2024jxm}.

Fig.~\ref{fig:4cases} shows the $(S/N)$ as a function of the coupling, in the four highlighted points of the parameter space in the left panel of fig.~\ref{fig:bound_s4}. The green point (bottom right panel) represents a region where the necessary coupling for the $(S/N)$ to reach unity is the smallest and there is further potential for enhancement with even higher coupling values. Additionally, it is never surpassed by the signal-to-noise ratio of the auto-correlation. In contrast, for the blue point (top left panel) the cross-correlation fails to always win over the auto-correlation and the $(S/N)$ achieves unity at much larger coupling values. The red point (top right panel) is within the excluded region of fig.~\ref{fig:bound_s4}, indeed its $(S/N)$ never reaches unity. Finally, the orange point (bottom right) permits a coupling value that elevates its signal-to-noise ratio to one. However, the signal-to-noise ratio of the auto-correlation surpasses this value even before it is reached. In all cases, we show the trend for the three CMB experiments under considerations, observing a noticeable decrease in the signal-to-noise ratio for less sensitive surveys, as anticipated\footnote{This decline ceases when the signal-to-noise ratio saturates, as explained by eqs.~\eqref{eq:cross_sat} and \eqref{eq:auto_sat}, where the noise term becomes negligible regardless of the target experiment.}.

We now briefly come back to the comparison with the results of ref.~\citep{Greco:2024oie}, which delineates the contours for the ALP-parameters reproducing the isotropic birefringence angle of $(0.3^\circ\pm0.11^\circ)$, constrained by \citep{Diego-Palazuelos:2022dsq}. We observe that, for the majority of points in their analysis, our parameter space offers a plethora of points with a signal-to-noise ratio exceeding unity. This is due to the rotation angle being degenerate with respect to the axion-photon coupling and initial misalignment angle (see eq.~\eqref{eq:alpha_inst}). In contrast, the cross-correlation of birefringence and galaxies displays a non trivial dependence on these parameters (see eq.~\eqref{eq:alphaG}) and can effectively break this degeneracy. For a more comprehensive comparison we refer to appendix \ref{app:A}. Additionally, it is important to highlight the phase degeneracy of the isotropic measurement \citep{Naokawa:2024xhn}, as potential experiments cannot distinguish between $\alpha_0$ and $\alpha_n = \alpha_0 + n \cdot 180^\circ$. Consequently, we expect the findings presented in ref.~\citep{Greco:2024oie} to extend to a broader parameter space. To clarify how our results compare with isotropic birefringence, we include a vertical line in each panel of fig.~\ref{fig:4cases} corresponding to the coupling value that yields $\alpha_0 = 0.3^\circ$ and its first phase-shift $\alpha_1 = 180.3^\circ$. These values can be directly calculated from eq.~\eqref{eq:alpha_inst} using $g_{\phi\gamma}^n = 2\,\alpha_n/(\phi_0 - \phi_s)$. They tend to decrease as the mass decreases or the initial misalignment increases. Axions with smaller masses begin evolving later, resulting in a larger difference between the field’s present value and that at last scattering. Similarly, larger initial misalignment angles amplify the field’s evolution. In all cases shown in fig.~\ref{fig:4cases}, $\alpha_0$ lies before the region of interest, where the signal-to-noise ratio exceeds unity. However, $\alpha_1$ aligns well with this region, and further phase-shifts would correspond to an even higher $(S/N)$.

In conclusion, for $m_\phi\lesssim10^{-32}\,\si{\electronvolt}$ (shifting towards $10^{-30.5}\,\si{\electronvolt}$ as $|\theta_i|$ approaches 0) and $|\theta_i|\gtrsim\pi/4$, we find an allowed value of the axion-photon coupling that results in a signal-to-noise ratio for the cross-correlation exceeding both unity and that of the auto-correlation. Moreover, this promising region of parameter space corresponds to axion-fields whose evolution commences around the epoch of reionization. This underlines this approach as an outstanding resource to constrain the origin of birefringence and axion-like physics. Furthermore, future data analyses employing this approach have the potential to establish stringent bounds on the axion-photon coupling, as we predict viable values to leave within $g_{\phi\gamma}\in[10^{-16},\;10^{-12}]\,\si{\giga\electronvolt^{-1}}$.

\section{Conclusion}
\label{sec:conc}
In this paper we have discussed, for the first time, the cross-correlation between anisotropic cosmic birefringence and galaxy number counts as a probe of axion-like physics. Parity violating terms in the electromagnetic Lagrangian, such as ALPs coupling to photons, induce the "in vacuo" rotation of the plane of linearly polarized waves (i.e. cosmic birefringence). The magnitude of this effect is heavily influenced by the path traveled by photons. Therefore, the cosmic microwave background, being the oldest light remnant observable today and having traveled vast distances to reach us, serves as a crucial tool for cosmology. It provides valuable information about the early universe, offering insights into fundamental cosmological parameters and the processes that shaped cosmic evolution. Moreover, the characteristics of the underlying axion model significantly impact the predicted rotation angle. We focus on ultralight axion-like particles, whose field oscillations begin at relatively late cosmological times. This aligns with the period when the density term from galaxy number counts tends to peak, underscoring the valuable insights that can be gleaned from this approach. We leverage EDE models from the string axiverse and explore their implications on this phenomenon.

We have computed this novel cross-correlation by following the background and perturbation evolution of the axion-like field, within a properly modified version of the Boltzmann code {\tt CLASS}. In this process, we ensured that the background field consistently behaves as a spectator field, meaning its energy contribution remains negligible compared to standard $\Lambda$CDM components and does not influence the Hubble expansion rate. Our analysis reveals significant variations in the correlation's amplitude across the parameter space characterizing the axion-like particle ($10^{-33}\,\si{\electronvolt}\le m_\phi\le10^{-28}\,\si{\electronvolt}$, $-\pi<\theta_i\le\pi$). We find a non negligible signal especially for large and intermediate angular scales ($\ell<100$). Additionally, we investigated how the axion-photon coupling dictates the amplitude of the birefringence angle and consequently influences the observed signal.

With the intent of finding when the cross-correlation is measurable by observations, we have explored the ALP-parameter space over a $100\times100$ grid on $m_\phi\,$-$\,\theta_i$ and computed the signal-to-noise ratio, exploiting the \emph{Euclid} survey's forecast specifications combined with those of future CMB experiments (LiteBIRD, Simons Observatory and CMB-S4). In particular, for CMB-S4 - the best performing of the three - we find the needed value of the coupling to reach unity in the $(S/N)$, conservatively keeping it below current bounds ($g_{\phi\gamma}\le10^{-12}\,\si{\giga\electronvolt^{-1}}$). We additionally calculate the signal-to-noise ratio for the auto-correlation of birefringence, demonstrating that for $m_\phi\lesssim10^{-32}\,\si{\electronvolt}$ and $|\theta_i|\gtrsim\pi/4$, there is not only a "relatively small" coupling value where the signal-to-noise ratio of the cross-correlation reaches unity, but the cross-correlation also provides more discernible information in terms of measurability compared to the auto-correlation alone. Additionally, these axion-parameters are linked to a late-time evolution of the field, occurring around reionization. This alignment with the epoch and the scales of interest of the galaxy counterpart serves to enhance the signal. This elucidation underscores the effectiveness of this approach in unraveling the quest to understand the origin of birefringence.

Given the upcoming release of DR1 from the \emph{Euclid} survey and the substantial influx of polarization data anticipated over the next decade, this approach stands as an exceptional tool for testing axion-like physics and investigating anisotropic cosmic birefringence. Moreover, existing polarization data from the {\it Planck} satellite and the ACT collaboration, in conjunction with current galaxy surveys, offer potential avenues for estimating this innovative cross-correlation, which we defer to future studies.

\acknowledgments
We thank Eiichiro Komatsu, Toshiya Namikawa, Luca Pagano, Zachary Slepian, Patrick Stengel, Sunny Vagnozzi and Luca Visinelli for valuable discussions. We also convey our gratitude towards the Institute for Fundamental Physics of the Universe (IFPU), that hosted the focus week "Parity violation through CMB observations"\footnote{\url{https://www.ifpu.it/focus-week-2024-05-27/}}, where the many insights enabled us to finalize our work. We acknowledge financial support from the INFN InDark initiative and from the COSMOS network (www.cosmosnet.it) through the ASI (Italian Space Agency) Grants 2016-24-H.0, 2016-24-H.1-2018 and 2020-9-HH.0.
We acknowledge the use of CINECA HPC resources from the InDark project in the framework of the INFN-CINECA agreement. 
SA and ML acknowledge participation in the COST Action CA21106 "COSMIC WISPers", supported by COST (European Cooperation in Science and Technology). SA acknowledges financial support by “Bando Giovani anno 2023 per progetti di ricerca finanziati con il contributo 5x1000 anno 2021”.
NB, AG and PN acknowledge support by the MUR PRIN2022 Project “BROWSEPOL: Beyond standaRd mOdel With coSmic microwavE background POLarization”-2022EJNZ53 financed by the European Union - Next Generation EU.
We express our appreciation to John Legend for the poignant title "Conversations in the Dark" drawn from the emotive lyrics that resonate deeply with our scientific exploration.

\appendix
\section{Impact of the potential recipe}
\label{app:A}

As discussed in section \ref{subsec:potential}, we focused our work on early dark energy models from the string axiverse, characterized by the potential (same as eq.~\eqref{eq:potential}, although repeated here for convenience)
\be\label{eq:A1}
    V(\phi) = m_\phi^2\,f_a^2\left[1-\cos{\frac{\phi}{f_a}}\right]^n \,,
\ee
where the exponent $n$ is fixed at 3, in order to align with previous works \citep{Poulin:2018dzj, Kamionkowski:2014zda}. This parameter governs the dynamics of the field, where for $n=1$, we recover the classical axion potential resembling pressureless behavior, and for $n=2$, the field evolves akin to a radiation-like fluid. However, when the exponent is equal to or less than 2, a resonance emerges in the evolution of perturbations (see eq.~\eqref{eq:EoMpert}) due to the frequency of the background oscillations matching that of perturbations, close to a specific Fourier scale $k$. This phenomenon can be thoroughly analyzed using Floquet theory \citep{ASENS_1883_2_12__47_0}, a branch of ordinary differential equations theory. When applied to EDE potentials, it leads to an exponentially growing solution of the EoM of perturbations \citep{Amin:2011hu,Smith:2023fob}. Hence, for these values of $n$ a non-perturbative approach should be adopted and we leave it to future works.
\begin{figure}[t]
\centering
\includegraphics[width=0.95\hsize]{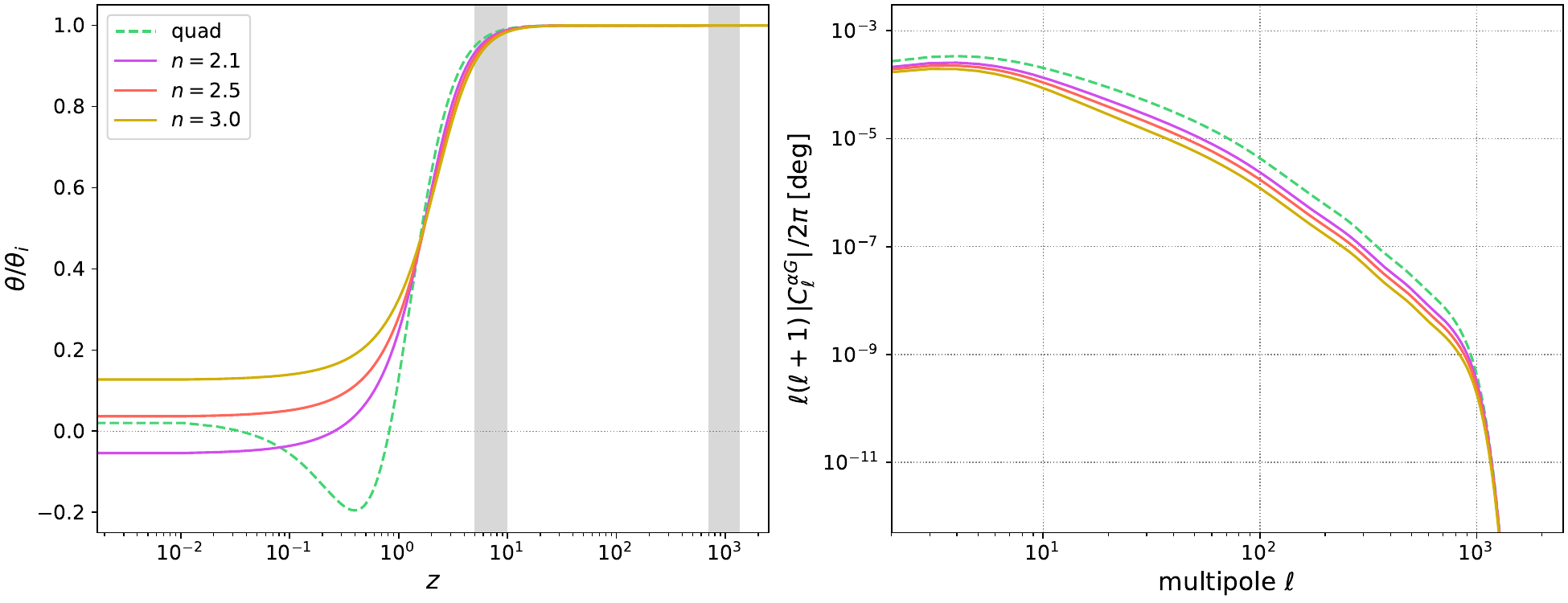}
\caption{(\emph{Left}): solution to the background equation of motion (eq.~\eqref{eq:EOMbkg}), as a function of redshift, for three values of the exponent for an EDE potential, as shown in eq.~\eqref{eq:A1}, and for a quadratic potential (see eq.~\eqref{eq:A2}). The two gray-shaded regions represent the windows of reionization and recombination, respectively. (\emph{Right}): cross-correlation angular power spectrum between anisotropic birefringence and the spatial distribution of galaxies, corresponding to these four potential-settings. The results have been computed with the following ALP-parameters: $m_\phi=10^{-32}\,\si{\electronvolt}$, $\theta_i=\pi/2$, $g_{\phi\gamma}=\SI{2e-14}{\giga\electronvolt^{-1}}$.}
\label{fig:varpot}
\end{figure}

Within the perturbative-allowed region ($n>2$), we performed our analysis for different values of the exponent, as well as for a separate model, accounting for a simple quadratic potential
\be\label{eq:A2}
    V(\phi)=\frac{1}{2}\,m_\phi^2\,\phi^2 \;.
\ee 
Let us notice, that the latter corresponds precisely to the expansion, near the minimum, of eq.~\eqref{eq:A1}, when $n=1$. Consequently, we do expect a non-perturbative behavior at some scales. Specifically, we are unable to investigate regions of the parameter space where $m_\phi\gtrsim \SI{2.2e-32}{\electronvolt}$, due to the onset of oscillations early enough to trigger exponential growth in the perturbation evolution. 

Fig.~\ref{fig:varpot} illustrates, in the left panel, the solution to the background EoM for three distinct values of the exponent (within an EDE model described by eq.~\eqref{eq:A1}), as well as for the aforementioned quadratic potential. For all cases, we considered the following ALP-parameters: $m_\phi=10^{-32}\,\si{\electronvolt}$, $\theta_i=\pi/2$, $g_{\phi\gamma}=\SI{2e-14}{\giga\electronvolt^{-1}}$. The field uniformly begins its evolution around the epoch of reionization, as the main parameter is the mass and its interplay with the Hubble expansion rate. Nevertheless, a larger value of the exponent reflects on the field diluting faster (see section \ref{subsec:potential}). In the case of the quadratic potential, we anticipate a quicker onset of oscillations, which aligns with the $n=1$ scenario. As a result, the cross-correlation signal, depicted in the right panel, is most pronounced for the latter case, gradually decreasing for larger exponent values.
\begin{figure}[t]
\centering
\includegraphics[width=0.95\hsize]{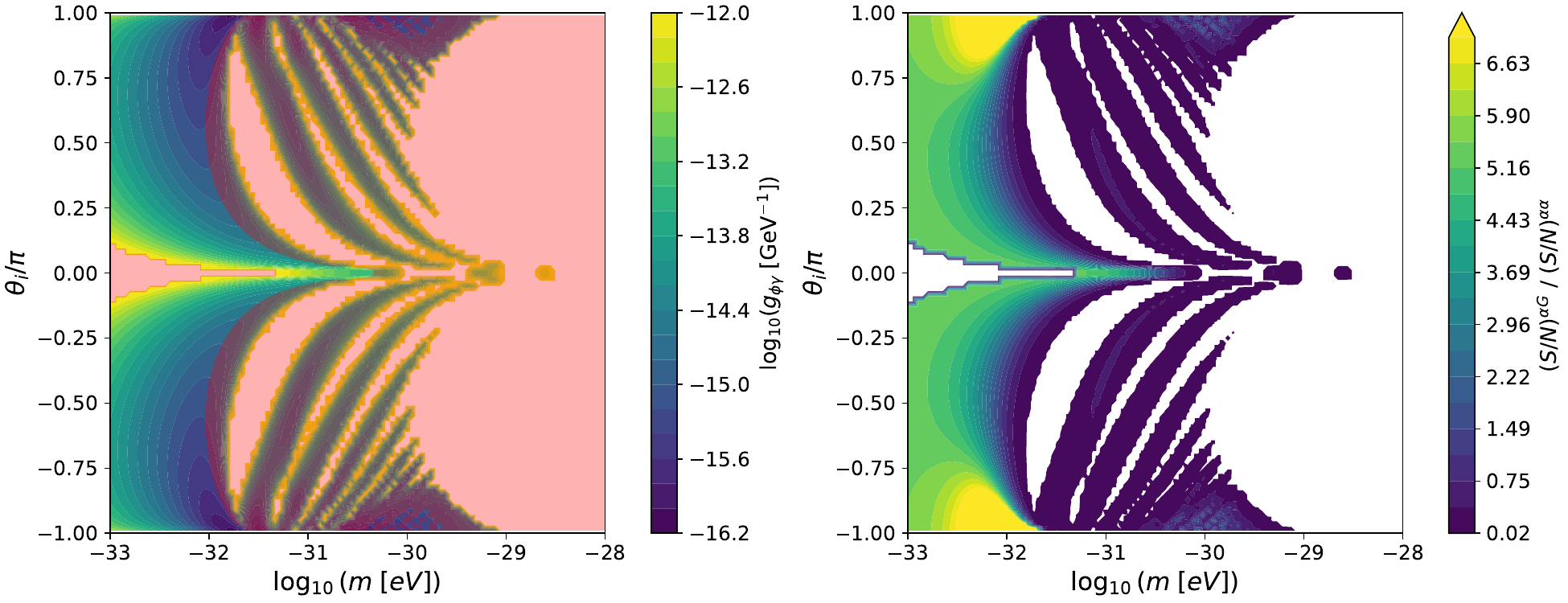}
\caption{Same as fig.~\ref{fig:bound_s4}, with $n=2.1$ as exponent of the EDE potential of eq.~\eqref{eq:A1}.}
\label{fig:bound_n2.1}
\end{figure}

\begin{figure}[t]
\centering
\includegraphics[width=0.95\hsize]{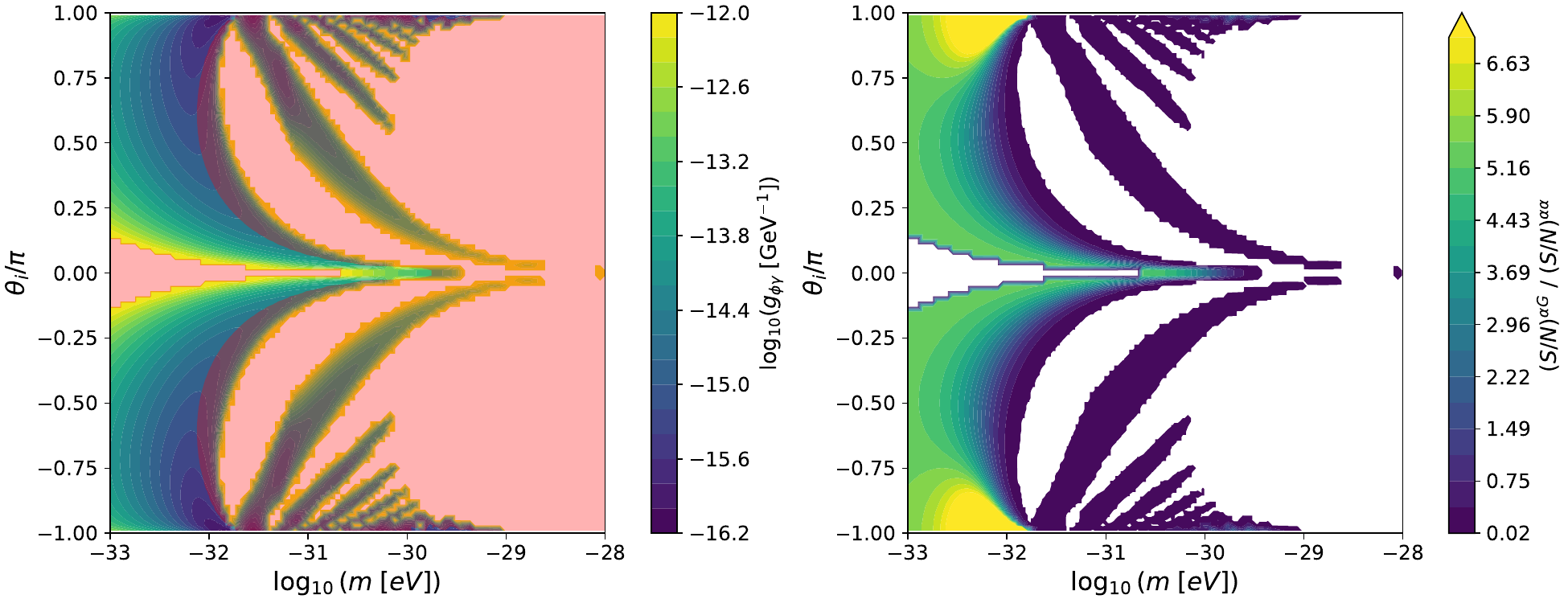}
\caption{Same as fig.~\ref{fig:bound_s4}, with $n=2.5$ as exponent of the EDE potential of eq.~\eqref{eq:A1}.}
\label{fig:bound_n2.5}
\end{figure}

\begin{figure}[t]
\centering
\includegraphics[width=0.95\hsize]{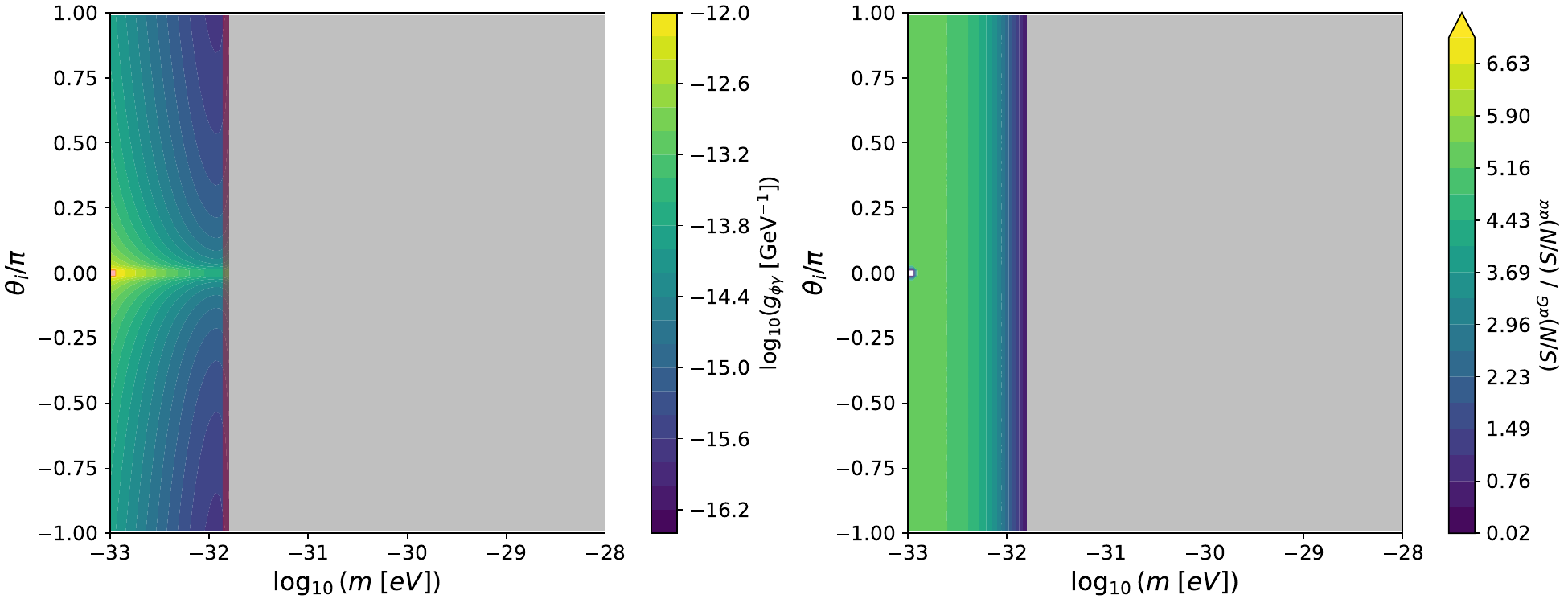}
\caption{Same as fig.~\ref{fig:bound_s4}, with a quadratic potential as a reference model. Masses beyond $10^{-32}\,\si{\electronvolt}$ are excluded, due to the perturbative approach not being applicable here.}
\label{fig:bound_quad}
\end{figure}

In terms of detectability, we show the contour plots for the axion-photon coupling needed to achieve unity in the signal-to-noise ratio, as detailed in section \ref{sec:res}, in figs.~\ref{fig:bound_n2.1}, \ref{fig:bound_n2.5} and \ref{fig:bound_quad}. The first two pertain to an EDE model with $n=2.1$ and $n=2.5$, respectively. The outcomes are akin to those obtained for $n=3$, although the coupling values across the parameter space are diminished, and new features emerge at higher masses as the exponent decreases. Indeed, smaller values of $n$ correspond to a fluid that dilutes more slowly, resulting in larger cross-correlation signals. 

The constrained region for a quadratic potential is depicted in fig.~\ref{fig:bound_quad} and displays a different behavior with respect to the previous cases. Nevertheless, the conclusions remain consistent, bringing about a preferred region of the parameter space for masses around $m_\phi=10^{-32}\,\si{\electronvolt}$ and large initial misalignment angles. It's important to note that we excluded all points with $m_\phi>\SI{2.2e-32}{\electronvolt}$ as the perturbative approach is not reliable in this domain (see discussion above). In particular, the latter case can be compared to the region constrained by \citep{Greco:2024oie}, where the isotropic measurement of the birefringence angle \citep{Diego-Palazuelos:2022dsq} is reproduced. Upon examining their fig.~(1a) in the parameter space $m_\phi$-$g_{\phi\gamma}\theta_i$, we observe how each point in their constrained region corresponds to numerous points in ours, as the cross-correlation is not degenerate in the couple $g_{\phi\gamma}$, $\theta_i$. As a consequence, this study not only introduces a novel tool for constraining ALP-parameters and understanding the nature of cosmic birefringence but also aligns its results with current measurements of the isotropic birefringence angle.

\section{Impact of tomography}
\label{app:B}
Our work shows, for the first time, the usefulness of cross-correlating anisotropic cosmic birefringence with galaxies, as both are sourced by the metric perturbations. Through this innovative approach, we are able to constrain the region of $m_\phi\,$-$\,\theta_i$ parameter space where axion-photon coupling values, within available bounds, allow for a measurable signal in future experiments. Our analysis adopts a non-tomographic approach, focusing on a single redshift bin within the Euclid prescriptions \citep{Euclid:2021icp}. Nonetheless, it is crucial to estimate the impact of a full-tomographic strategy on our findings. Thus, we extend eq.~\eqref{eq:S/Ncross} to incorporate the complete covariance across the 10 redshift bins of \emph{Euclid} \citep{Euclid:2019clj}:\footnote{The edges of the 10 equi-populated bins are:\\
$z_{\rm edge}=\{0.001,\,0.42,\,0.56,\,0.68,\,0.79,\,0.90,\,1.02,\,1.15,\,1.32,\,1.58,\,2.50\}$.}
\be\label{eq:S/Ntomo}
    (S/N)^{\alpha G}_{\rm tomo} = \sqrt{\sum_{\ell=2}^{\ell_{\rm max}}\left(C_\ell^{\alpha G}\right)^\intercal {\rm Cov}_{\alpha G}^{-1}\left(C_\ell^{\alpha G}\right)} \;,
\ee
where $C_\ell^{\alpha G}$ is the vector containing the cross-correlation signal for each bin and ${\rm Cov}_{\alpha G}^{-1}$ the associated inverse covariance matrix, computed following eq.~\eqref{eq:deltacross}.

\begin{figure}[t]
\centering
\includegraphics[width=0.49\hsize]{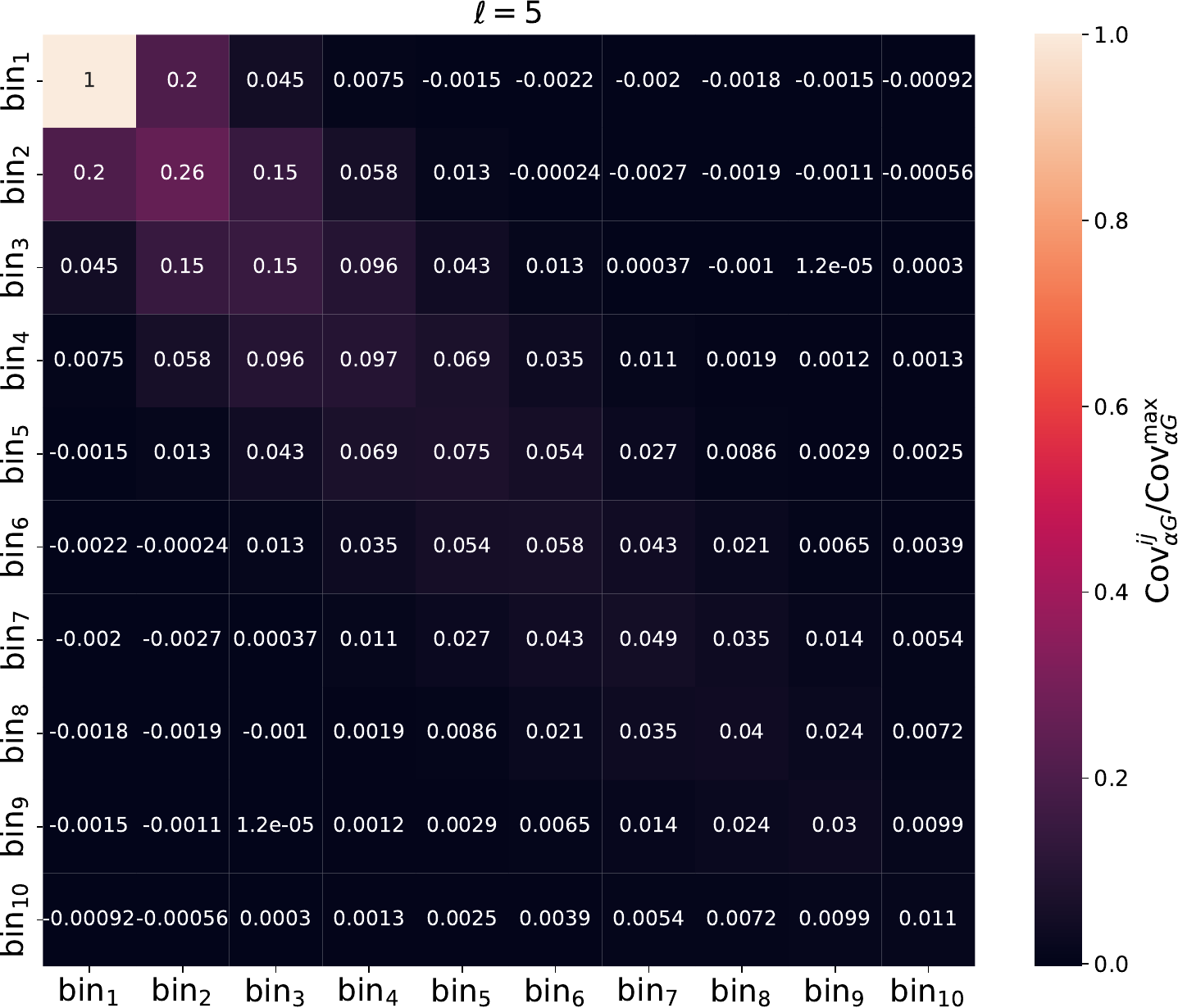}
\includegraphics[width=0.49\hsize]{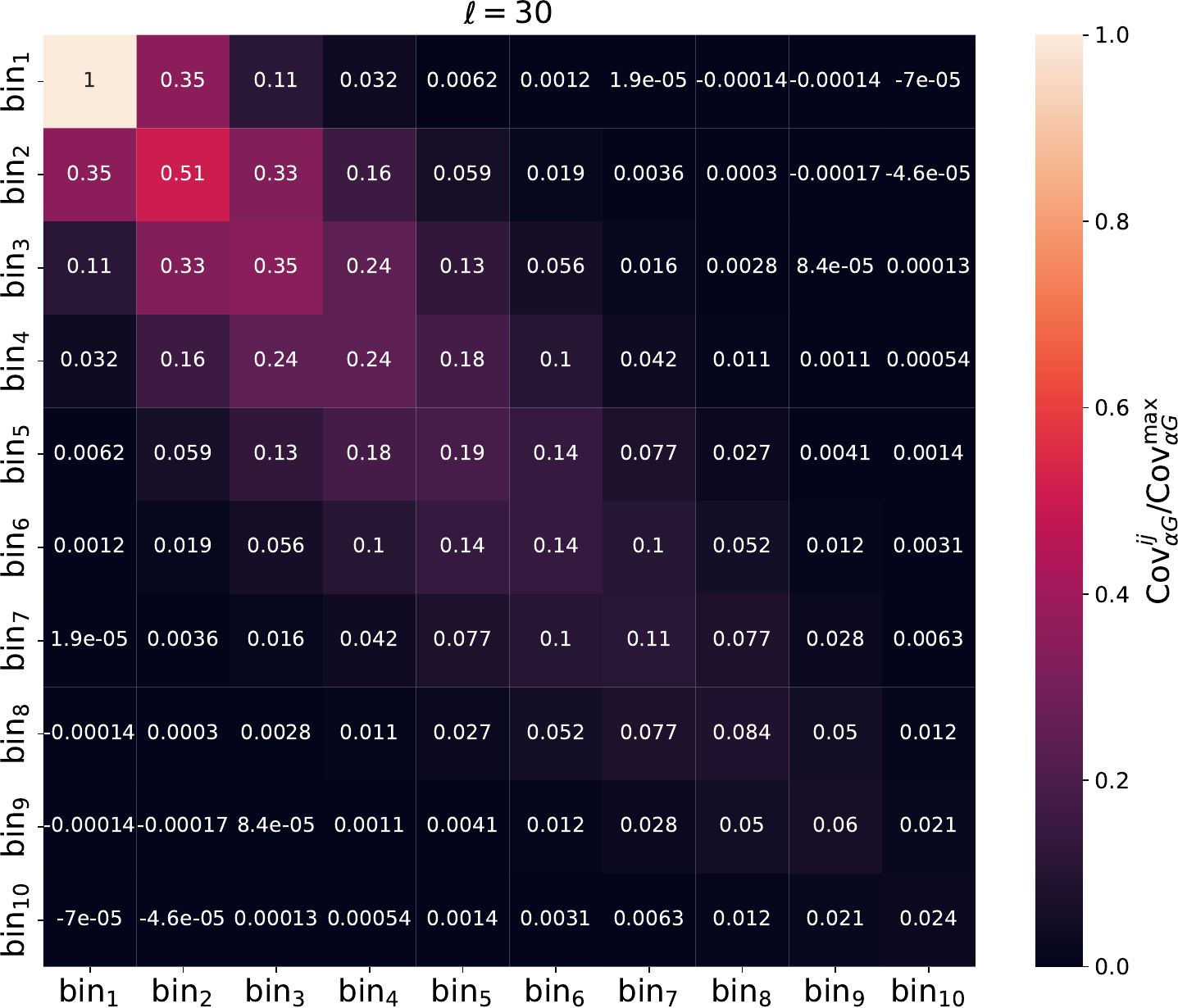}
\caption{Covariance matrix for the birefringence-galaxy cross-correlation in the tomographic case, across the 10 redshift bins of \emph{Euclid} \citep{Euclid:2019clj} (higher $i$'s correspond to higher redshifts) up to $z=2.5$. The matrix is normalized to its maximum value and shown for $m_\phi=10^{-32}\,\si{\electronvolt}$, $\theta_i/\pi=0.999$ and $g_{\phi\gamma}=\SI{2e-14}{\giga\electronvolt^{-1}}$. We show the result for $\ell=5$ in the left panel and $\ell=30$ in the right panel.}
\label{fig:covmat}
\end{figure}
\begin{figure}[t]
\centering
\includegraphics[width=0.65\hsize]{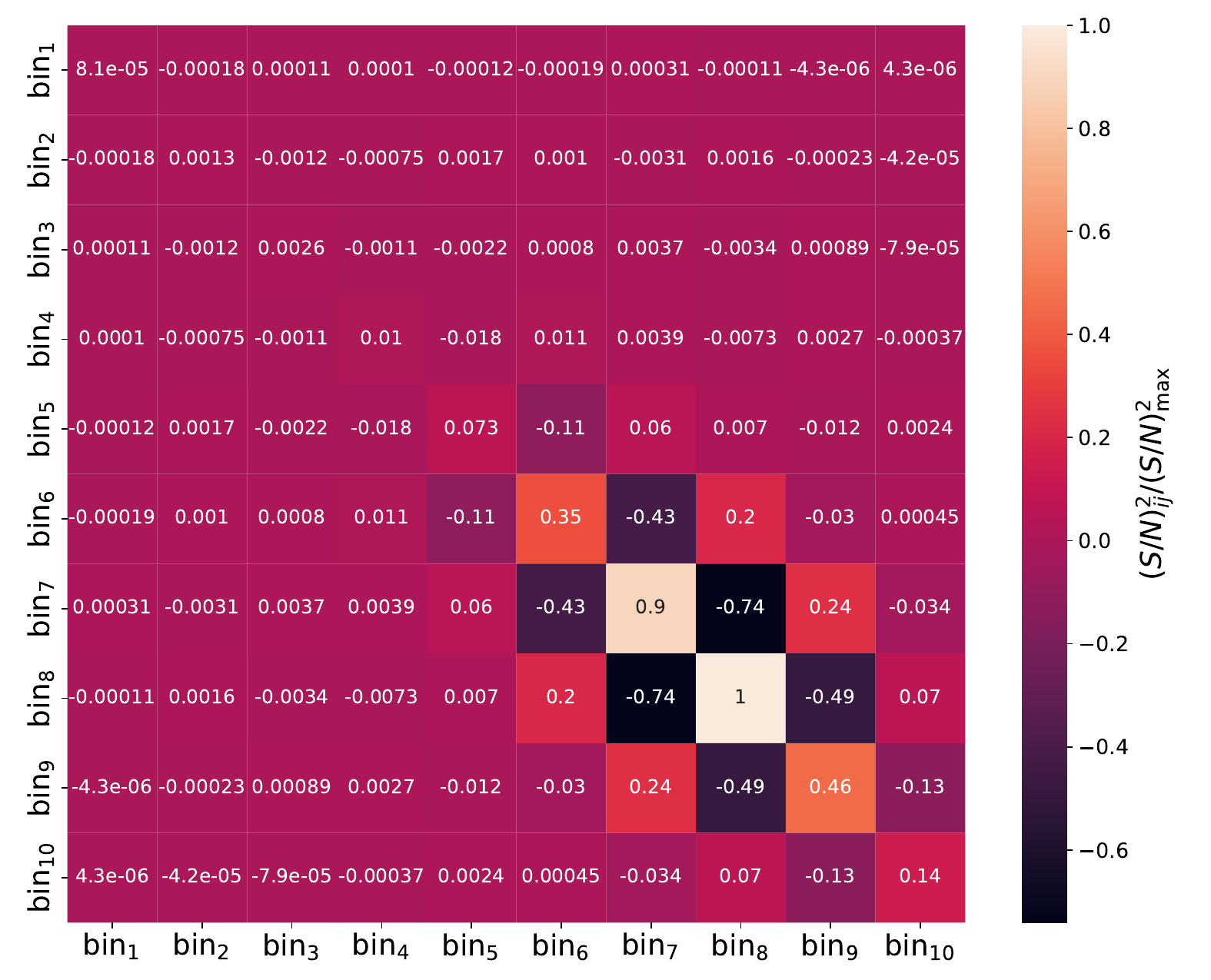}
\caption{Contributions to the $(S/N)^{\alpha G}$ from each bin$_i\,$-$\,$bin$_j$ correlation, across the 10 redshift bins of \emph{Euclid} \citep{Euclid:2019clj} (higher $i$'s correspond to higher redshifts) up to $z=2.5$. The result is normalized to its maximum value and shown for $m_\phi=10^{-32}\,\si{\electronvolt}$, $\theta_i/\pi=0.999$ and $g_{\phi\gamma}=\SI{2e-14}{\giga\electronvolt^{-1}}$. Let us notice that negative terms arise from possible anti-correlations between different redshift bins, whilst the total sum exploited in eq.~\eqref{eq:S/Ntomo} is, of course, positive definite.}
\label{fig:covmat2}
\end{figure}

Fig.~\ref{fig:covmat} illustrates the covariance matrix for $m_\phi=10^{-32}\,\si{\electronvolt}$, $\theta_i/\pi=0.999$ and $g_{\phi\gamma}=\SI{2e-14}{\giga\electronvolt^{-1}}$, for two different multipoles ($\ell=5$ in the left panel and $\ell=30$ in the right). The $i$-th redshift bin corresponds to higher redshifts as $i$ increases. Examining eq.~\eqref{eq:deltacross}, we anticipate the diagonal terms to be dominant as the galaxy shot noise only contributes here. Nevertheless, it's notable how the first off-diagonal terms are non-negligible and possibly important in the full-tomographic computation of the signal-to-noise ratios. Moreover, the covariance exhibits great variability with multipoles, with the latter behavior being more pronounced as the multipole increases. Similarly, in fig.~\ref{fig:covmat2} we calculate, following eq.~\eqref{eq:S/Ntomo}, the contributions to the $(S/N)$ from each bin$_i\,$-$\,$bin$_j$ correlation\footnote{Despite the signal-to-noise ratio being positive definite overall, it can display negative contributions from any off-diagonal correlation (in case of an anti-correlation between bin$_i$ and bin$_j$).}, for the same point in the ALP-parameter space. This result not only reaffirms how the first off-diagonal terms are not negligible, but also shows how each bin$_i\,$-$\,$bin$_j$ couple contributes to the total signal-to-noise ratio.

\begin{figure}[t]
\centering
\includegraphics[width=0.95\hsize]{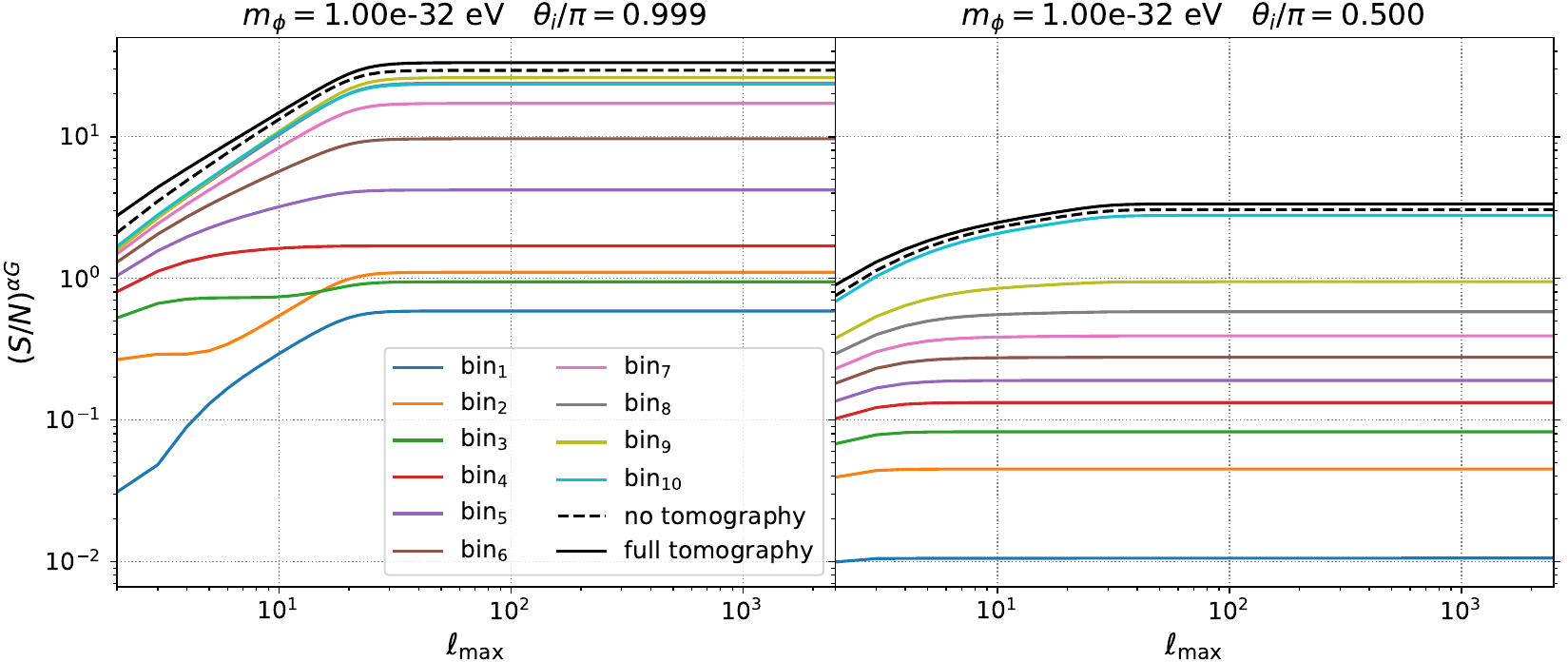}
\caption{Signal-to-noise ratios of the birefringence-galaxy cross-correlation as a function of the maximum multipole $\ell_{\rm max}$ for two points in the $m_\phi\,$-$\,\theta_i$ parameter space. The result is shown for the ten redshift bins of \emph{Euclid} \citep{Euclid:2019clj} (higher $i$'s correspond to higher redshifts), as well as for the full tomographic approach of eq.~\eqref{eq:S/Ntomo} (solid). We compare with the non-tomographic approach (dashed) followed in the main text. Tomography leads up to a 20\% improvement. The axion-photon coupling is set to $\SI{2e-14}{\giga\electronvolt^{-1}}$.}
\label{fig:snr_ell_tomo}
\end{figure}
Finally, exploiting eq.~\eqref{eq:S/Ntomo} we compute the total $(S/N)$ for each of the ten redshift bins\footnote{We consider solely the diagonal terms in the covariance matrix for the $(S/N)$ computation in each of the ten redshift bins.}, as well as the cumulative outcome from employing the full tomographic method, for two points within the ALP-parameter space. We compare them with the non-tomographic approach in fig.~\ref{fig:snr_ell_tomo}. It becomes apparent that the majority of information arises from the latter bins, corresponding to higher redshifts. This pattern is attributed to the anticipation of larger signals where the galaxy kernel peaks. Moreover, tomography allows for slightly better results, with the signal-to-noise ratio improving up to the 20\% level. Given that our primary aim is to introduce the cross-correlation between birefringence and galaxy number counts as a valuable probe of ALPs, we defer the exploration of expanded analyses using a full tomographic approach to future studies.

\section{Energy density of the axion}
\label{app:C}
\begin{figure}[t]
\centering
\includegraphics[width=0.95\hsize]{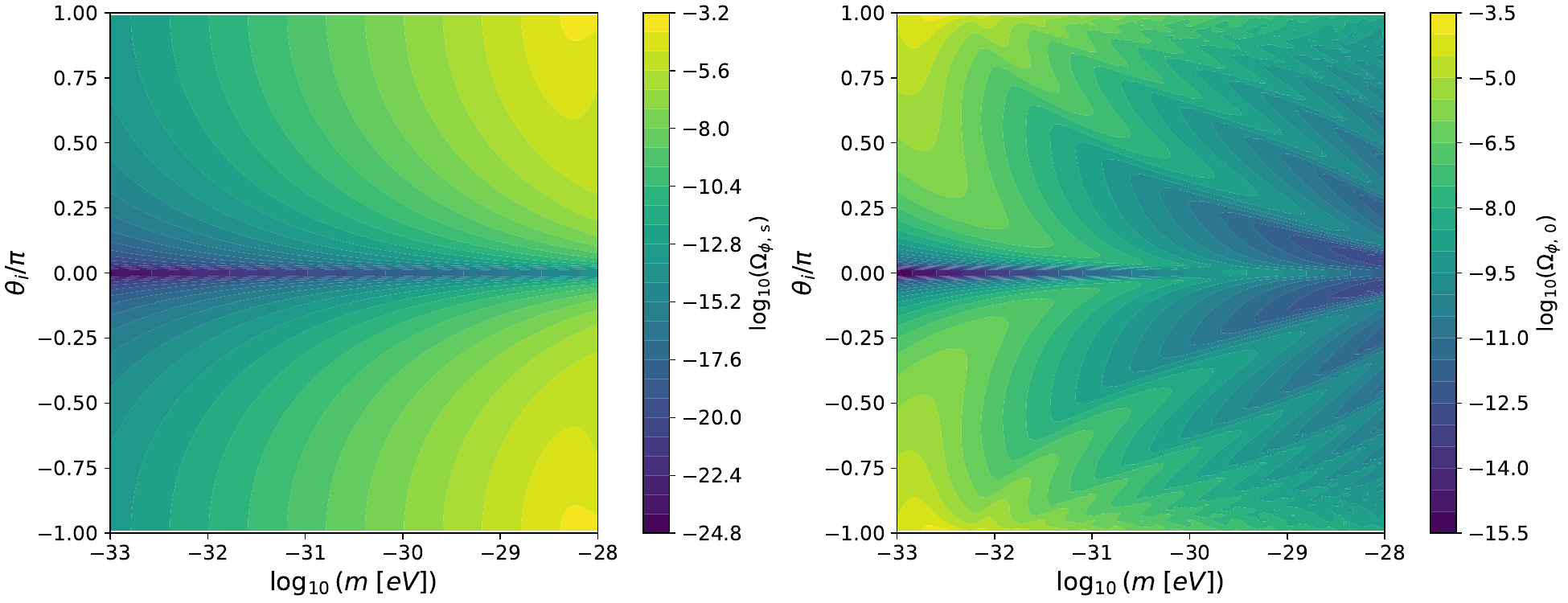}
\caption{Axion energy density parameter for each mass-initial misalignment couple of our parameter space at last scattering $\Omega_{\phi,\,s}$ (\emph{Left}) and today $\Omega_{\phi,\,0}$ (\emph{Right}).}
\label{fig:energy}
\end{figure}

As discussed in the main text, our analysis is conducted under the assumption that the spectator field approximation remains valid across the entire parameter space. This ensures that the energy density, as expressed in eq.~\eqref{eq:energy}, remains negligible compared to the total energy budget of the universe at all times. Consequently, the axion field does not influence the expansion history or the current value of the Hubble parameter. However, it is insightful to examine how the energy density behaves throughout the $m_\phi\,$-$\,\theta_i$ parameter space. In fig.~\ref{fig:energy}, we present the axion energy density parameter $\Omega_\phi = \rho_\phi / \rho_c$ (where $\rho_c$ is the universe’s critical density) at the time of last scattering and today, shown in the left and right panels, respectively. These values are, at most, three orders of magnitude below unity and follow a predictable trend. Larger masses cause the evolution to begin closer to the recombination period, thus contributing more to the energy budget here. Conversely, smaller masses lead to later evolution, resulting in a greater contribution today. Additionally, higher initial misalignment angles yield larger contributions by amplifying the field’s overall evolution.

\bibliographystyle{JHEP}
\bibliography{ref}

\end{document}